\documentclass[
 aip,
 amsmath,amssymb,
 reprint,%
]
{revtex4-2}

\usepackage{graphicx}% Include figure files
\usepackage{dcolumn}% Align table columns on decimal point
\usepackage{bm}% bold math
\usepackage{floatrow}
\usepackage[caption=false]{subfig}
\usepackage[english]{babel}

\usepackage{xcolor}
\usepackage{hyperref}
\hypersetup{
    colorlinks=true,
    linkcolor=blue,
    filecolor=magenta,      
    urlcolor=cyan,
}
\usepackage{soul} % for \hl
\usepackage[normalem]{ulem}

\newcommand{\heinz}[1]{{\color{blue}#1}}

\newcommand{\beq}{\begin{equation}}
\newcommand{\eeq}{\end{equation}}
\newcommand{\bea}{\begin{eqnarray}}
\newcommand{\eea}{\end{eqnarray}}
\newcommand{\p}{\partial}
\newcommand{\mb}{\mathbf}

\newcommand{\apjl}{Astrophysical Journal Letters}
\newcommand{\mnras}{Monthly Notices of the Royal Astronomical Society} 
\newcommand{\aap}{Astronomy \& Astrophysics}
\newcommand{\jgr}{Journal of Geophysics Research}

\begin{document}

%\preprint{}

\title{Formation and Evolution of Coherent Structures in 3D Strongly Turbulent Magnetized Plasmas}% Force line breaks with \\
%\thanks{Footnote to title of article.}

\author{Loukas Vlahos}  
\homepage{https://orcid.org/0000-0002-8700-4172, vlahos@astro.auth.gr, corresponding author}
\author{Heinz Isliker}
\homepage{https://orcid.org/0000-0001-9782-2294, isliker@astro.auth.gr}
\affiliation{Department of Physics,
              Aristotle University, 54124 Thessaloniki, Greece}
%\author{}%
 %\email{Second.Author@institution.edu.}
%\affiliation{ Department of Physics,
              %Aristotle University, 54124 Thessaloniki, Greece}%\\This line break forced with \textbackslash\textbackslash
%

%\author{Anastasios Anastasiadis}
 %\homepage{http://www.Second.institution.edu/~Charlie.Author.}
%\affiliation{Institute for Astronomy, Astrophysics, Space Applications and Remote Sensing,\\
%National Observatory of Athens,
%GR-15236 Penteli, Greece}%\\This line break forced% with \\
%

\date{\today}% It is always \today, today,

\begin{abstract}
We review  the current  literature on the formation of Coherent Structures ({\bf CoSs}) in strongly turbulent 3D magnetized plasmas. CoSs  (Current Sheets ({\bf CS}), magnetic filaments, large amplitude magnetic disturbances, vortices, and shocklets) appear intermittently inside a turbulent plasma and are collectively the locus of magnetic energy transfer (dissipation) into particle kinetic energy,  leading to  heating and/or acceleration of the latter. CoSs and especially CSs are also evolving and fragmenting, becoming  locally the source of new clusters of CoSs. Strong turbulence can be generated by the nonlinear coupling of large amplitude  unstable plasma modes, by the explosive reorganization of large scale magnetic fields, or by the fragmentation of CoSs. A small fraction of CSs inside a strongly turbulent plasma will end up reconnecting. Magnetic Reconnection ({\bf MR}) is one of the potential forms of energy dissipation of a turbulent plasma. Analysing the evolution of  CSs and  MR in isolation from the surrounding CoSs and plasma flows may be convenient for 2D numerical studies, but it is far from a realistic modeling of 3D astrophysical, space and laboratory environments, where strong turbulence can be exited, as e.g.\ in the solar wind, the solar atmosphere, solar flares and  Coronal Mass Ejections (CMEs), large scale space and astrophysical shocks, the magnetosheath, the magnetotail, astrophysical jets, Edge Localized Modes (ELMs)  in confined laboratory plasmas (TOKAMAKS), etc. 
\end{abstract}

%Use showkeys class option if keyword
\maketitle
%\tableofcontents

\section{Introduction \label{Intro}}
Strong  turbulence is a complex nonlinear dynamic phenomenon, which has a great impact on the heating and acceleration of particles  in space and laboratory plasmas \cite{FrischU95, Biskamp03}.  Unfortunately, courses for the study of turbulence are little present in university graduate programs. As a result,  strong turbulence is also absent from the modeling of laboratory, astrophysical and  space  phenomena when they enter into a fully developed turbulent stage. The basic plasma physics courses at the universities start with the exploration of normal modes and linear instabilities. The nonlinear evolution of  unstable waves is analysed  with the use of the quasilinear approximation. In laboratory, space and astrophysical plasmas the "linear" phase of a normal mode has no meaning since the fluctuations grow in the presence of strong turbulence. The estimate of the growth time of the fluctuations in the presence of fully developed turbulence remains an open problem.  Recently, Prof.\ William H.\  Matthaeus   wrote a review article with the provocative title 
  {\it  ``Turbulence of space plasmas: Who needs it?" }\cite{MatthaeusREv2021}
to stress the  following fact: the scientific community avoids the use of strong turbulence in the interpretation of many  astrophysical or laboratory plasma phenomena. Most studies  treat the linear part of the evolution of a system very carefully, but when their models enter into the regime of fully developed turbulence, the intermittent appearance  of Coherent Structures (CoSs) and their multi-scale evolution fall beyond the ability  of their numerical tools to handle them with  present day computers. Therefore,  the interpretations of many 3D strongly turbulent space and laboratory phenomena remain unexplored. 

\subsubsection{Weak vs strong turbulence}

The study of turbulence can be  divided into ``weak''  or ``wave'' turbulence and ``strong'' turbulence. We define as ``weak'' or ``wave'' turbulence  the magnetic fluctuations resulting from the superposition of any spectrum of $N$ linear modes,  \begin{equation} \label{weak}
    \mb{b}(\mb{r},t)=\sum_{i=1}^{N} \mb{b}_{0i} e^{i(\mb{k}_i \cdot \mb{r} -\omega(\mb{k}_i)t+\phi_i)} ,
\end{equation}
where $\mb{k}_i$ is the wave vector, $\omega(\mb{k}_i)$ the  dispersion relation derived through linearization, $\mb{b}_{0i}$ the amplitude and $\phi_i$ the random phase of the the weakly damped/amplified wave   mode $i$.
This is a correct representation of a physical system if its unstable  fluctuations have very small amplitude (i.e.\ for magnetized plasmas the fluctuations of the magnetic field $\mb{b}$ are very weak, $|\mb{b}|<<|\mb{B_0}|$, where $\mb{B_0}$ is the ambient magnetic field of the plasma). 

In weak (wave) turbulence there is spectral transfer of energy through the resonant three wave interaction, analyzed with the use of the quasi linear theory and the prescribed energy dissipation at the small wave lengths \cite{Vedenov63, Galtier09}.

Many references to "turbulence" in the current literature refer to "weak" turbulence, since its mathematical description is relatively easy, and with the use of the quasilinear approximation one can estimate the transport properties of the particles \cite{Vedenov63}. The regime where the unstable waves reach large amplitudes ($|\mb{b}| \geq |\mb{B_0}|$ ) is  called ``strong'' turbulence if the turbulence is nearly isotropic \cite{Goldreich95, Perez08}.   The non-linear evolution of the magnetic disturbances controls the energy transfer between the different scales and the particles. The most important characteristic of strong turbulence, which is not present in weak turbulence,  is the intermittent appearance of CoSs. In this review, we focus on the 3D aspects of strong turbulence, and especially we address the question how  CoSs are formed and evolve. The role CoSs play in the  fast heating and acceleration of particles during explosive phenomena in astrophysical and laboratory plasmas is currently an open problem and cannot be addressed properly before having a good understanding of the statistical properties of the multi-scale evolution of CoSs \citep{Vlahos19}.

\subsubsection{Intermittency and coherent structures}

The magnetic energy and its dissipation in strongly turbulent systems is concentrated  into intermittently appearing and disappearing CoSs. However,  it is not well understood how CoSs are formed and distributed inside a turbulent volume.  Also, the CoSs play a crucial role in  how the dissipated energy is partitioned, and they operate on different scales \cite{Wan16}. An important assumption in fluid turbulence is that the energy injected at large scales is     transferred to smaller and smaller scales by non-linear processes, where it is dissipated when it reaches the kinetic scales \cite{Pezzi21}. This process is known as the turbulent {\bf energy cascade}. If CoSs of all sizes are distributed inside a strongly turbulent volume, the dissipation of energy via heating and acceleration of particles is distributed over all scales, and not only the small (kinetic) scales. 3D numerical  simulations generally  show   turbulence  to  be  quite  intermittent, and  scaling  models  have  been 
developed to incorporate this effect\cite{Karimabadi2013c}.  Particles interacting with  a collection of CoSs (CSs, shocks, large amplitude magnetic perturbations)  on all scales gain or lose energy as they travel through the turbulent volume before escaping \cite{Sioulas22c}. It is not apparent how the energy is dissipated on the different scales, and a detailed study is needed to clarify this point.

As we are going to show in this review, turbulence  can  generate, among other types of CoSs, current sheets (CSs)  at  all scales, and it has been proposed that a fraction  of the CSs undergo magnetic reconnection  as part of 
their  dissipation  process \cite{Carbone90}. It is important to stress that CSs are only one of the many types of CoSs appearing in strongly turbulent magnetised plasma.  The other types include large amplitude magnetic disturbances, magnetic filaments, vortices, shocklets, and tangential discontinuities \cite{Tsurutani79, Neugebauer84} (the list is still not complete).

\subsubsection{2D vs 3D coherent structures and magnetic reconnection}

\begin{figure}[ht!]
%\sidesubfloat[]
{\includegraphics[width=0.5\columnwidth]{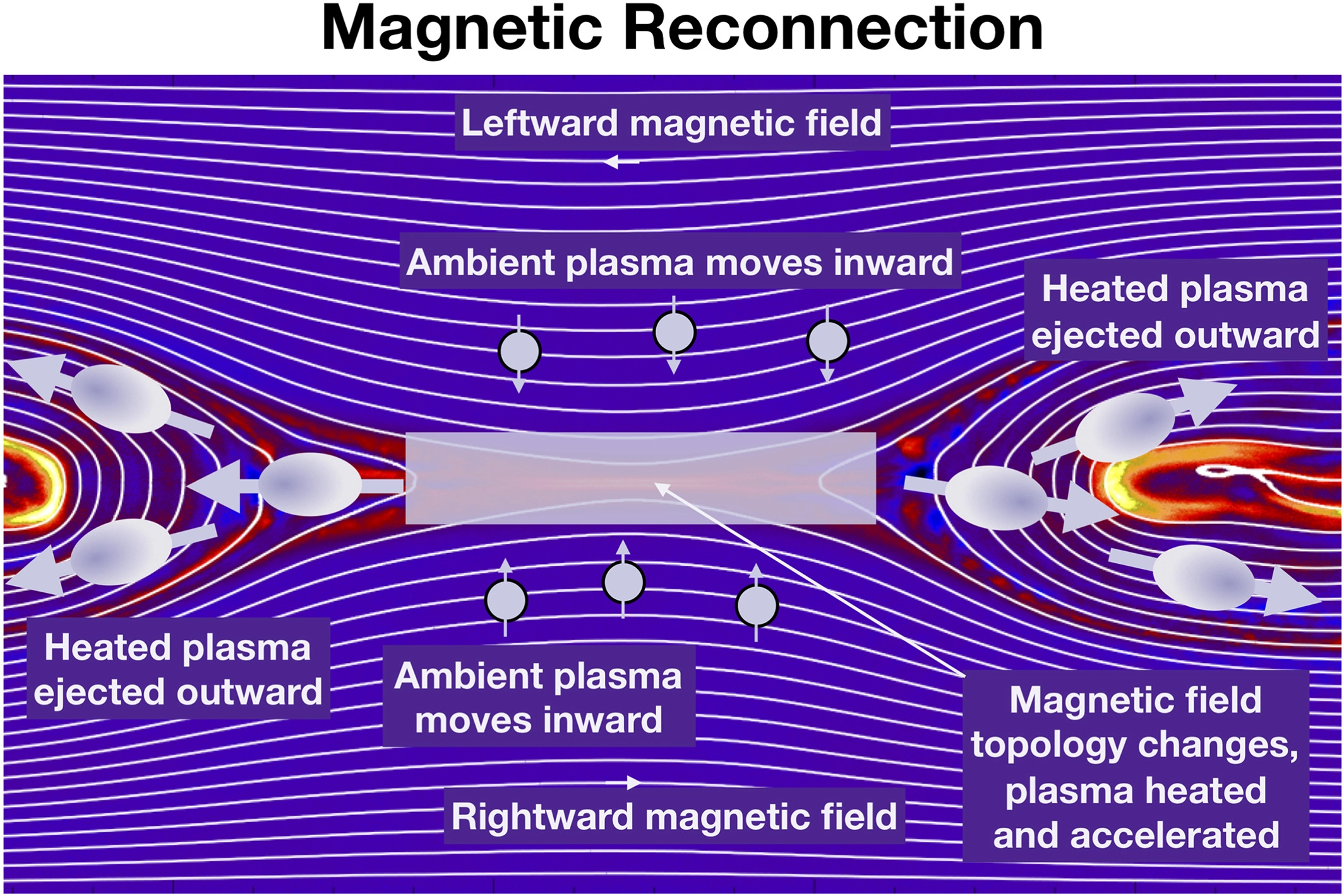}%
		\label{CSIsolated}}\\
		%\sidesubfloat[]
		{\includegraphics[width=0.6\columnwidth]{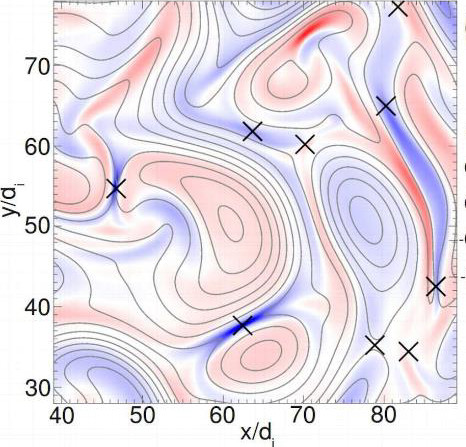}%
		\label{CSturb}}\\
		%\sidesubfloat[]
		{\includegraphics[width=0.6\columnwidth]{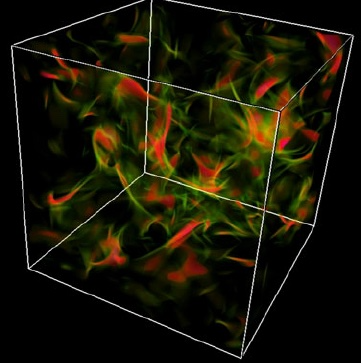}%
		\label{MHD_t}}\\
	\caption{
		%\protect\subref{CSIsolated} 
		(a) Most numerical studies start with a CS already present in the middle of a 2D periodic simulation box, the driver for the reconnecting CS is either a laminar or ``weakly'' turbulent flow. Reproduced with permission from Hesse and Cassak, Journal of Geophysical Research (Space Physics) {\bf 125}, e25935 (2020). Copyright 2020 Wiley.  
		%\protect\subref{CSturb} 
		(b) The formation of  CSs in 2D strong turbulence    is driven by large scale magnetic fluctuations and/or  other coherent structures intermittently formed in their vicinity. The CSs are never alone and isolated. The reconnecting sites are marked with crosses and represent a small fraction of the CSs formed. Reproduced with permission from Servidio et al., Physical Review Letters, {\bf 102} 115003 (2009). Copyright 2009 APS.  
		%\protect\subref{MHD_t} 
		(c) A snapshot of the spatial distribution of the   electric current  density in 3D MHD strong turbulence. The formation of 3D CSs is  a fundamental aspect of 3D strong turbulence. Reproduced with permission of Minnini et al., Journal of Plasma Physics, {\bf 73}, 377-401 (2007) Copyright 2007 Cambridge University Press. 			}\label{CS2}
			\end{figure}

Almost all studies on magnetic reconnection so far start with a current sheet already formed in the middle  of a 2D periodic simulation box (see Fig. \ref{CS2}) \cite{Parker57, Petschek64, Zwebel16, Loureiro2016, Hesse2020}. The problem of CS formation in 3D magnetic topologies  and the characteristics of  convective flows that drive the reconnection process remained outside the scope of these studies for many years. The evolution of an isolated CS in the presence of weak turbulence in the incoming flows has been analysed recently \cite{Lazarian99, Kowal09, Vishniac12, Kowal2012}. The characteristics of the convective flows, which,  among others,
 act as the driver for the reconnection process and are responsible for the breaking of the initial CS and the formation of plasmoids or secondary CSs in 2D numerical simulations, have been analyzed extensively \cite{Comisso15, Zwebel16, Loureiro2016}.

With the presence of a second CS inside the simulation box, the evolution of magnetic reconnection leads to a multi island environment as well \cite{Coppi6, Matthaeus86, Drake06, Hoshino12, Adhikari20}. When starting a simulation with strong turbulence, the CSs appear intermittently at random places  inside the 2D periodic simulation box, and their evolution depends on the complex drivers and the CoSs surrounding the reconnecting CSs \cite{Servidio09} (see Fig. \ref{CS2}), and finally, in 3D simulations, the  formation of a CS driven by a strongly turbulent surrounding is shown in Fig. \ref{CS2} \cite{Mininni07}. The motion of CoSs also generates waves that are emitted into the ambient plasma. \cite{Karimabadi2013c} Therefore, the intermittent formation of CSs in a 3D strongly turbulent plasma departs radically from the evolution of an isolated CS in 2D, as shown in Fig. \ref{CS2}.

In the early 80's, the link between reconnection and turbulence has been established \cite{Matthaeus86},  and a few years later the link between turbulence and reconnection has also been analyzed \cite{Biskamp89}. Several recent reviews discuss the way how turbulence can become the host of reconnecting current sheets and how  reconnecting current sheets can drive turbulence \cite{Matthaeus11, Cargill12, Lazarian12, Karimabadi2014}. The link between shocks and turbulent reconnection has also been analyzed \cite{Karimabadi2014}.

Strong turbulence and CoSs in the solar atmosphere are driven by the convection zone, and the spontaneous formation of reconnecting and non-reconnecting CSs has been analyzed in several articles \cite{Parker83, Parker88, Galsgaard96, Galsgaard97a, Gasgaard97b, Einaudi21}.

Our review will focus on the formation and evolution of CoSs inside a 3D strongly turbulent magnetized plasma. In section II, we explore the way how strong turbulence generates CoSs, in section III we analyse the fragmentation and filamentation of a 3D large scale isolated CS, which eventually leads to the formation  of a cluster of CoSs and  strong turbulence. In section IV, we discuss the presence of strong turbulence and CoSs upstream and downstream of a shock.  In Section V, we explore the way how strong turbulence is driven by the convection zone, by emerging magnetic flux, or by unstable large scale magnetic structures in the solar atmosphere. In section VI,  we use  methods from complexity theory (e.g.\ Cellular Automata and Self Organized Criticality) to explore  the formation and evolution of CoSs (mainly CSs)    in a strongly turbulent magnetized plasma. In section VII, we summarise the main points of this review.

\section{Formation of coherent structures in strong  turbulence} \label{turrec}

 There are several ways to initiate strong turbulence in 2D and 3D numerical simulations \cite{Biskamp89,Dmitruk03, Arzner06,Servidio09,Servidio10,Servidio11,Zhdankin13,Valentini14,Karimabadi2014,Cerri17,Isliker17a,Comisso19,Rueda21}.   In this section, we follow the approach used initially by   Dmitruk et al. \cite{Dmitruk04} and later by Arzner at al \cite{Arzner06}, Zhdankin et al. \cite{Zhdankin13} and Isliker et al. \cite{Isliker17a}. In these articles, the authors   did not set up a specific geometry of a reconnection environment or prescribe a collection of waves \cite{Arzner04} as turbulence model, but allow the MHD equations themselves to build
naturally correlated field structures (which are turbulent, not random) and
coherent regions of intense current densities (current filaments or CSs).  In this approach, turbulence is freely evolving and ultimately decaying.

The 3D, resistive, compressible and normalized MHD equations used in \citet{Isliker17a} are
\beq
\p_t \rho = -\nabla \cdot \mathbf{p}
\eeq
\beq
\p_t \mathbf{p} =
- \mathbf{\nabla}  \cdot
\left( \mathbf{p} \mathbf{u} - \mathbf{B} \mathbf{B}\right)
-\nabla P - \nabla B^2/2
%+ \rho \bar{\nu} \nabla^2 \mb{v}
\eeq
\beq
\p_t \mathbf{B} =
-  \nabla \times \mathbf{E}
\eeq

\beq
\p_t (S\rho) = -\mathbf{\nabla} \cdot \left[S\rho \mathbf{u}\right]
\eeq
with $\rho$ the density, $\mathbf{p}$ the momentum density,
$\mathbf{u} = \mathbf{p}/\rho$,
$P$ the thermal pressure,
$\mathbf{B}$ the magnetic field,
\beq \mathbf{E}   = -  \mathbf{u}\times \mathbf{B} + \eta \mathbf{J}\eeq
the electric field,
$\mathbf{J} =  \mathbf{\nabla}\times\mathbf{B}$
the current density, $\eta$ the resistivity,
$S=P/\rho^\Gamma$ the entropy,
and $\Gamma=5/3$ the adiabatic index.

In \citet{Isliker17a}, the MHD equations are solved numerically in Cartesian coordinates with the pseudo-spectral method \cite{Boyd2001}, combined with the strong-stability-preserving Runge Kutta scheme \cite{Gottlirb98}, and by applying periodic boundary conditions to a grid of size $128\times 128\times 128$.  A fluctuating magnetic field $\mb{b}$ consist of a superposition of Alfv\'en waves, with a Kolmogorov type spectrum in Fourier space, together with
a constant background magnetic field $B_0$ in the
$z$-direction, so the  magnetic field used as initial condition is $\mb{B}=\mb{B}_0 +\mb{b}(x,y,z,t)$.  The mean value of the initial magnetic perturbation is $<b> = 0.6 B_0$, its standard deviation is $0.3 B_0$, and the maximum equals $2B_0$, so that indeed strong turbulence is considered. The initial velocity field is 0, and the initial pressure and energy are spatially constant.

\begin{figure}[h]
	\centering
	\includegraphics[width=0.70\columnwidth]{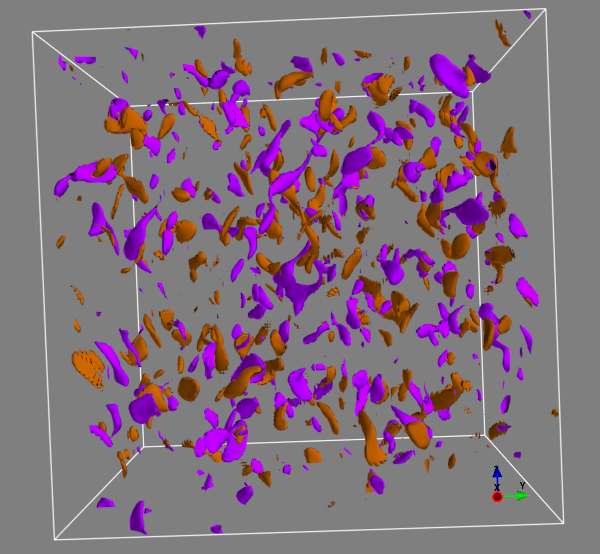}
	\caption{Iso-contours of the supercritical current density component $J_z$ (positive in brown negative in violet). Reproduced with permission from Isliker et al., Physical Review Letters, {\bf 119}, 045101 (2017), Copyright 2017 APS}
	\label{snapshot}
\end{figure}

The structure of  the $z$-component of the current density $J_z$  is shown in Fig. \ref{snapshot} ( the threshold value is chosen as the value above which the frequency distribution of the current density values deviates from Gaussian statistics, forming an exponential tail.).

For the CoSs to form, Isliker et al. \cite{Isliker17a} let the MHD equations evolve.
until the largest velocity component starts to exceed twice the Alvf\`en speed.
The magnetic Reynolds number at final time is $<|\mathbf{u}|>l/\eta = 3.5\times 10^3$  (being actually rather constant over time), with $l\approx 0.01$ a typical small scale eddy size,
and the ratio of the energy carried by the magnetic perturbation to the kinetic energy is $(0.5 <b^2>) /(0.5<\rho \mathbf{u}^2>) = 1.4$, which is a clear  indication that  they were dealing with strong turbulence.

\begin{figure}[ht!]
	\centering
		\includegraphics[width=0.70\columnwidth]{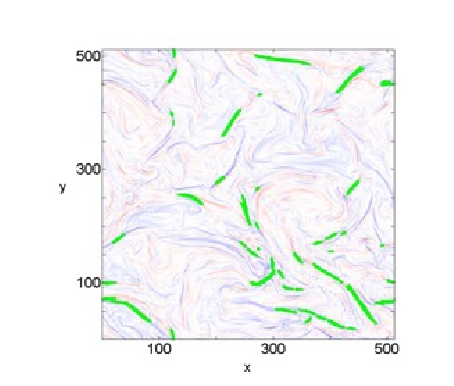}
	\caption{Current density in the cross section of the $x$-$y$-plane. Red indicates negative current, and blue indicates positive current. The presence of current sheets (in green color) throughout the volume is clearly visible.  Reproduced with permission from Zhdankin et al., The Astrophysical Journal, 771, 124 (2013), Copyright  2013 AAS.}
	\label{Distrcs}
\end{figure}

The overall picture demonstrates the spontaneous formation of CoSs, with the intermittent appearance and disappearance of CSs dominating the overall evolution of the strongly turbulent environment. This result resembles the 2D simulations of Biskamp and Walter \cite{Biskamp89} about thirty years ago. The perpendicular component of the current fluctuates rapidly but lacks the coherent structures shown in $J_z$. Similar results were obtained by Arzner et al. \cite{Arzner04,   Arzner06}, using strong Gaussian fields or a large eddy simulation scheme.

\begin{figure}[h]
%\sidesubfloat[]
{\includegraphics[width=0.8\columnwidth]{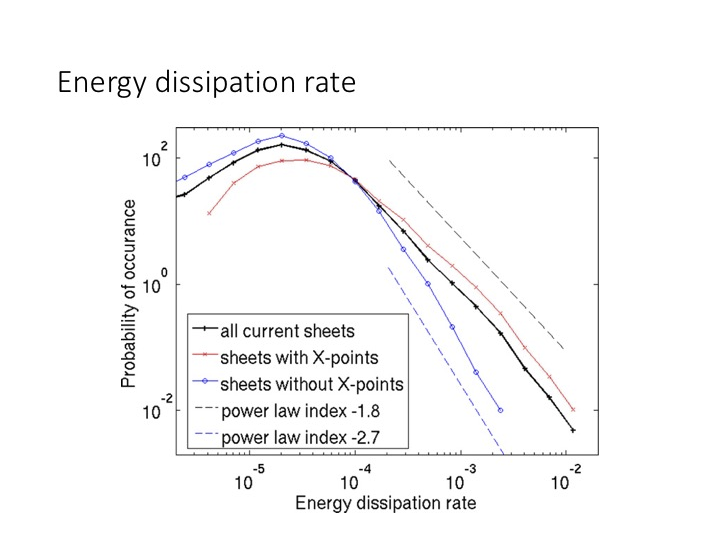}%
		\label{CSsta1}}\hfill\\
		%\sidesubfloat[]
		{\includegraphics[width=0.7\columnwidth]{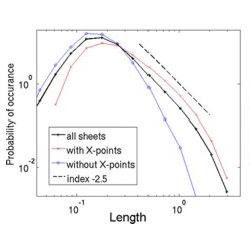}%
		\label{CSstat2}}\hfill
	\caption{(a) Probability distribution of the current sheet Ohmic dissipation rate. The distribution from all currents sheets shows a power law tail with index close to $-1.8$. (b) Probability distribution of the current sheet lengths. Reproduced with permission from Zhdankin et al., The Astrophysical Journal, {\bf 771}, 124 (2013), Copyright  2013 AAS.}
	\label{stat}
\end{figure}

It is of foremost importance to find ways to identify 3D CoSs inside a turbulent plasma and measure their statistical characteristics. Several algorithms have been proposed in order to identify and characterize the geometrical structures of CoSs in numerical simulations and observations\cite{Greco09,Servidio09,Servidio10,Servidio11,Uritsky10,Zhdankin13,Rossi15,DeGiorgio17,Fadanelli19,Perrone20,Sisti21,Sioulas22}.

\begin{figure}[h]
	\centering
	\includegraphics[width=0.60\columnwidth]{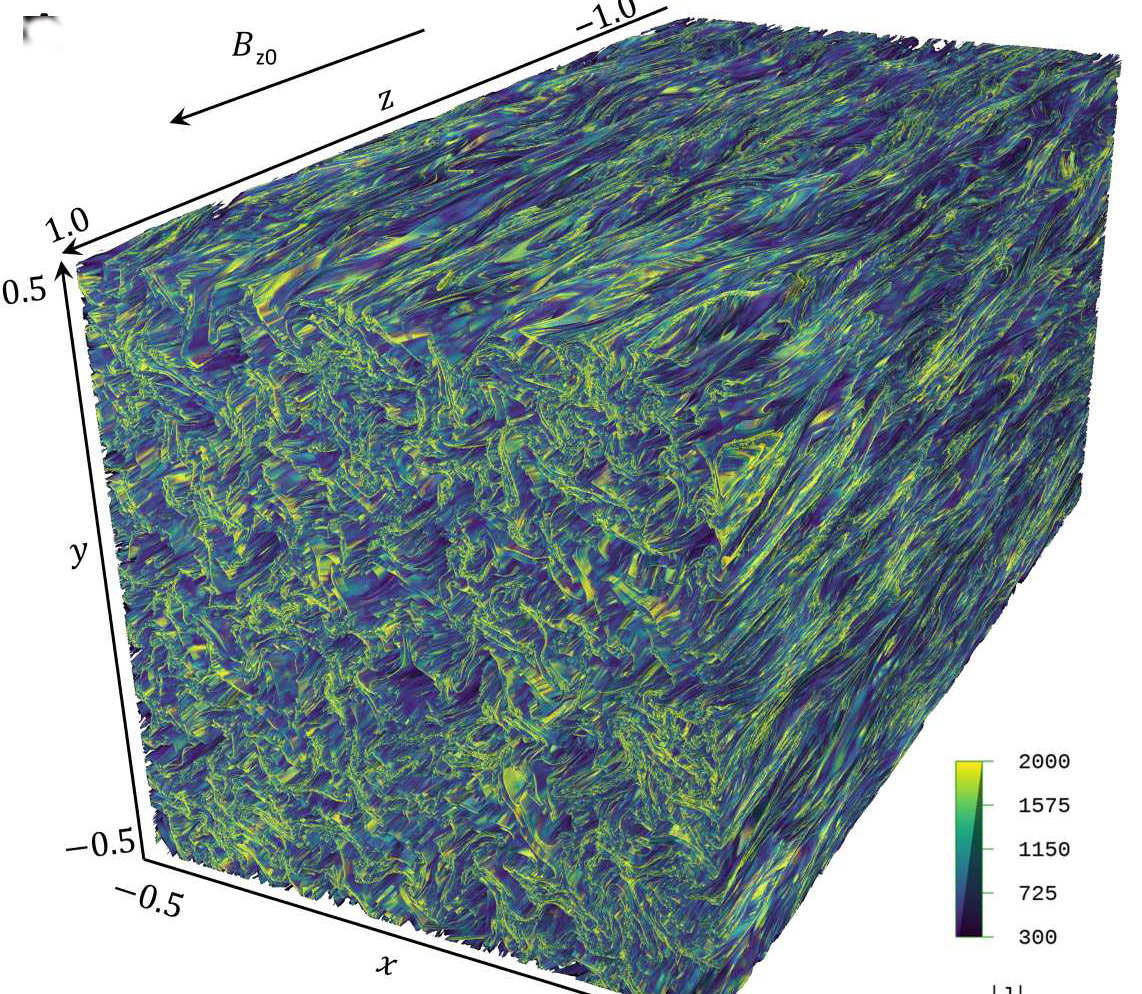}
	\caption{ Volume rendering of the current density $|J|$ in the entire domain at a stage when
turbulence is fully developed. A myriad of current sheets is evident in the plane perpendicular to the mean magnetic field $B_{z0} $  (for details of the simulation see Dong et al. \cite{Dong22}). Reproduced with permission from Dong et al., Sience Advances, {\bf 8}, 7627 (2022), Copyright  2022 AAAS.}
	\label{Dong1}
\end{figure}

Dong et al. \cite{Dong22} presented the world's largest, so far, 3D MHD turbulence simulation, using $\sim 200$ million Central Processing Unit (CPU) hours. In their analysis, a myriad of fine structures (CSs) is produced (see Fig. \ref{Dong1}),   

They initialized their simulation with uncorrelated, equipartitioned velocity and magnetic field fluctuations superimposed onto a strong mean magnetic field. Their main focus was on the properties of the reconnection-driven energy cascade, and they also discussed the role that the break up of the reconnecting CSs into smaller fragments plays in the energy transfer. The fragmentation of large scale CSs is the topic of the next section in this review. 

Comisso and Sironi \cite{Comisso22} showed that coherent structures (referring mainly to CSs) undergo fragmentation and reconnection in fully kinetic simulations of strong plasma turbulence. Their study proved  that reconnecting current sheets are a common feature of not only the MHD models but also of the more complete fully kinetic models.

Zhdankin et al. \cite{Zhdankin13} developed a framework for studying the statistical properties of CSs formed inside a magnetized plasma by using a 3D reduced MHD code. The distribution of the current fragmentation forming CSs in the $x$-$y$-plane is shown in Fig. \ref{Distrcs}. They were able to show that a large number of CSs do not contain reconnection sites, and likewise, many reconnection sites do not reside inside 3D CSs.

The most striking characteristic of the CSs formed spontaneously inside the strongly turbulent plasma is the probability distribution of the dissipated  energy, $
\varepsilon =\int \eta j^2 dV
$, and of the characteristic  lengths of the CSs, which are shown in Fig. \ref{stat}, as reported by Zhdankin et al \cite{Zhdankin13}. The techniques applied by Zhdankin et al. \cite{Zhdankin13} for the analysis of  their numerical simulations have initially been developed by Urisky et al. \cite{Uritsky10} . Recently, a number of attempts were made to extend the search for 3D CoSs to satellite data \cite{Fadanelli19,Sisti21}. 

\begin{figure}[h]
	\centering

	\includegraphics[width=0.70\columnwidth]{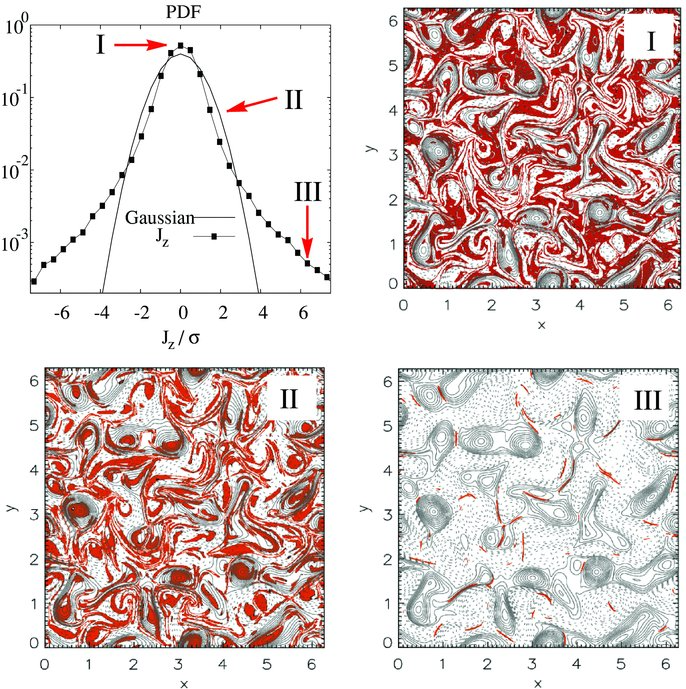}
	\caption{The PDF of the electric current density from 2D MHD simulations. Real space locations belonging to the regimes I, II and III are shown in the other three panels.Reproduced with permission from Greco et al. , The Astrophysical Journal Letters, {\bf 691}, L111 (2009), Copyright  2009 AAS.}
	\label{PDF}
\end{figure}

The distribution in real space of CoSs in turbulence influences their dynamics and dissipation characteristics through the complex interrelationship. In Fig. \ref{PDF}, the probability distribution function (PDF) of the electric current density is shown, and corresponding regions in real space are indicated \cite{Greco09}. The low values of the current density (region I) follow a supergaussian distribution and are related with the lanes between the islands. The intermediate values of the current density (region II) correspond to cores (filaments in 3D) and follow a subgaussian distribution, and finally the supergaussian tails of the distribution (region III) with the strongest current densities possibly represent the current sheets between interacting magnetic islands (filaments in 3D).

When confronted with a dataset that samples a turbulent  plasma system spatially, an important task is to find the subset of the data that corresponds to the underlying CoSs. In recent years, a plethora of methods has been suggested for the identification of intermittent structures and discontinuities in the magnetic field. These include the Phase Coherence Index method \citep{Hada03}
and the wavelet-based Local Intermittency Measure (LIM) \citep{Bruno99}. A simple and well-studied method that has been effectively used in the past for the study of intermittent turbulence and the identification of CoSs, both in simulations \citep{Servidio11} and observations \citep{Chasapis18, Chhiber20}, is the Partial Variance of Increments (PVI) method. The advantage of the PVI method is that it provides an easy-to-implement tool that measures the sharpness of a signal relative to the neighborhood of a point. For a lag $\tau$, the normalized  PVI at time $t$ is defined as
\begin{equation}
    PVI( t , \tau) \ = \  \frac{| \Delta\textbf{B}( t, \tau)|}{\sqrt{\langle | \Delta\textbf{B}( t,\tau)|^{2}\rangle}},
\end{equation}
where $| \Delta\textbf{B}( t, \tau)| \ = \ |\textbf{B}(t \ + \ \tau) \ -  \ \textbf{B}(t)| $ is the magnitude of the magnetic field vector increments, and $\langle...\rangle$ denotes the average over a window that is a multiple of the estimated correlation time.  PVI is a threshold method, so to proceed with the analysis, one imposes a threshold ${\theta}$ on the PVI and selects portions in which $PVI > {\theta}$.  Greco et al. \citep{Greco18} have shown that increments with $PVI > 3$  lie in the ``heavy tails'' observed in the distribution of increments and can thus be associated with Non-Gaussian structures (see Fig. \ref{PDF}). By increasing the threshold value $\theta$, one can thus identify the most intense magnetic field discontinuities like CSs and reconnection sites. Finally, note that the method is insensitive to the mechanism that generates the coherent structures and it is more general and could be applied to 3D structures as well with appropriate modifications. This means that the PVI can be implemented for the identification of any form of sharp gradients in the  magnetic field. A more comprehensive review of PVI, as well as a comparison with the aforementioned methods, appropriate for identifying discontinuities, can be found in Greco et al. \cite{Greco18}.

\begin{figure}[h]
	\centering
	\includegraphics[width=0.50\columnwidth]{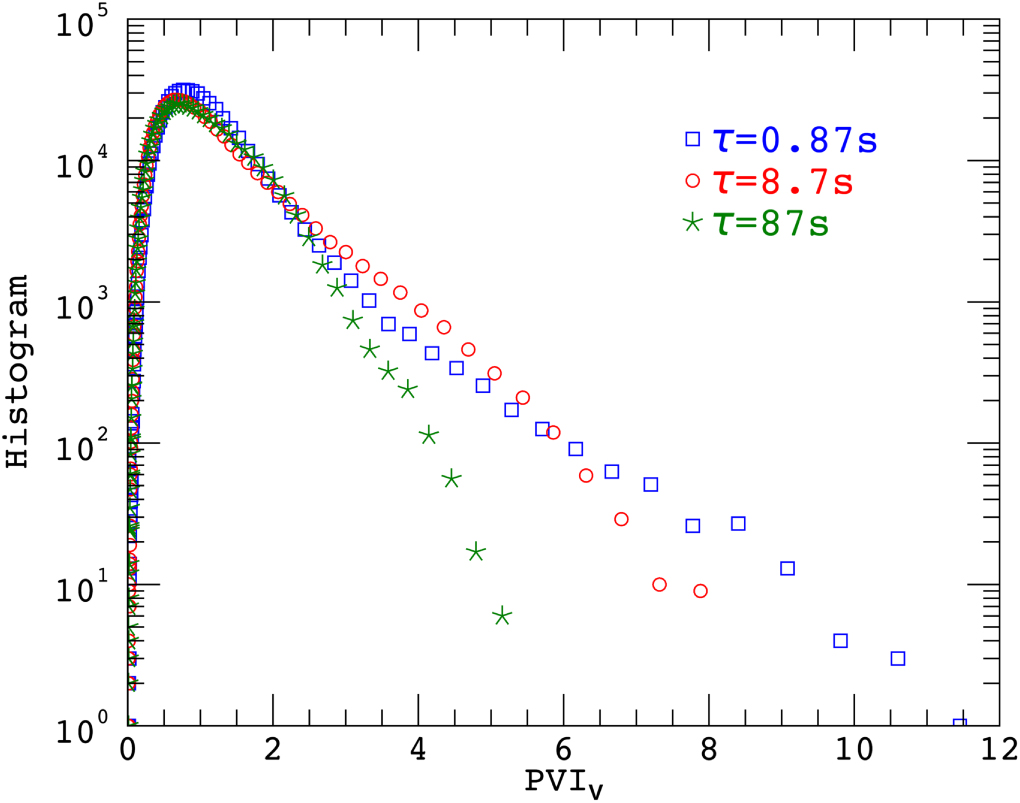}
	\caption{Histograms (frequency of occurrence, or number of counts) of PVI values for different lags $\tau$. Note the elevated likelihood of large PVI values at shorter lags, which is indicative of enhanced small-scale intermittency, typical of non-Gaussian processes and turbulence. Reproduced with permission from Chhiber et al., The Astrophysical Journal Supplement, {\bf 246}, 31 (2020), Copyright  2020 AAS.}
	\label{PVIHist}
\end{figure}

The histogram of the frequency of occurrence of PVI values in the solar wind data suggests that the most probable value of PVI is about 0.5 (see Fig. \ref{PVIHist}), where the majority of nonintermittent events takes place \citep{Chhiber20}. A large number of non-Gaussian ($3<PVI<6$) events appear  in the histogram in Fig. \ref{PVIHist}. The   percentage of possibly reconnecting events  ($PVI>8$) drops dramatically \cite{Chhiber20,Sioulas22}, suggesting, as we have stressed earlier \citep{Zhdankin13, MatthaeusREv2021}, that only a small fraction of the detected CSs are actually reconnecting. 

The techniques to explore the formation and evolution of CoSs at different scales in strongly turbulent magnetized plasma are in their infancy, and novel techniques are needed to understand   the statistical properties  of CoSs. We just mention that Jiang et al. \cite{Jiang21a} used convolutional neural networks to model strong  hydrodynamic turbulence and to follow the formation and evolution of CoSs. 

The main focus in most of the studies reported so far was on the presence  of intense CSs and especially reconnecting CSs inside the turbulent medium. Only recently the analysis was extended to other types of CoSs \cite{Perrone20}, e.g.\ vortex-like structures, wave packets, and Alfvenic fluctuations.

Karimabadi et al. \cite{Karimabadi2014} stress the fact that the motion of CoSs generates waves that are emitted into the ambient plasma in the form of highly oblique compressional  Alfven modes, as well as large amplitude magnetic disturbances.  This indicates that strong turbulence will in general consist of CoSs and waves, therefore "weak" and strong turbulence  co-exist in the multiscale evolution of a strongly turbulent plasma.
 Gro\v selj et al. \cite{Groselj19} explore the very important question of the relative importance of coherent structures and waves in strongly turbulent plasma, utilizing high-resolution observational and simulation data. They investigate the nature of waves and structures emerging in a weakly collisional, turbulent kinetic plasma. Their observational results are based on in situ solar wind measurements from the Cluster and the MMS spacecraft, and the simulation results are obtained from an externally driven, three-dimensional fully kinetic simulation. Using a set of novel diagnostics described in their article in detail, they show that both the large-amplitude structures and the lower-amplitude background fluctuations preserve linear features of kinetic Alfven waves to order unity. This quantitative evidence suggests that kinetic turbulence cannot be described as a mixture of mutually exclusive waves and structures, but may instead be pictured as an ensemble of localized, anisotropic wave packets or "eddies" of varying amplitudes, which preserve certain linear wave properties during their nonlinear evolution. This important finding and its role in the energy dissipation in strong turbulence, as well as its role in particle heating, has not been evaluated properly till now.

Large scale magnetic disturbances and CoSs in fully developed turbulence exhibit a monofractal or multi-fractal structure, both in space and astrophysical plasma \cite{Tu95, Shivamoggi97, Biskamp00, Leonardis13, Schaffner15, Isliker19}. This information is very important for analysing the interaction of particles with CoSs \cite{Sioulas20b}.

As we have already stressed, CSs are not the only coherent structures that appear in MHD turbulence. One also finds current cores, vorticity concentrations, and density structures \cite{Karimabadi13PP, Greco12, Servidio14}.  An informative example of neighboring coherent structures is found in the phenomenon of magnetic reconnection. It involves the formation of current sheets or filaments that become the organizational focus of the reconnection process as a whole.  This subtle interplay between coherent structures of different types (magnetic filaments, vortices, small-scale shocks, large amplitude magnetic disturbances) is indicative of the physics that controls the lower end of the MHD/fluid  cascade. These CoSs are likely the dominant loci of dissipative processes, not only in fluid models but also in kinetic approaches to plasmas \cite{MatthaeusREv2021}.

Concerning laboratory plasmas in tokamak devices, we note that turbulence appears predominantly at different spatial and temporal scales throughout the plasma
\cite{Martines02,Galassi17,Galassi19}.
Microscopic turbulence dominates in the plasma core, and it limits the steepness of the plasma temperature and density profiles \cite{Garbet04}. In the plasma edge region, microscopic and fluid turbulence can also be present in a dominant way when the plasma is in the so-called low confinement regime (L-mode). 
On the other hand, strong suppression of turbulence leads to the formation of a transport barrier and gives rise to the high confinement regime (H-mode), which exhibits a pedestal (extending over the edge transport barrier) that increases the core pressure \cite{Wagner82,Burrell97}.

Edge localized modes (ELMs) are violent and transient MHD instabilities that repeatedly take place in tokamak H-mode plasmas and are caused by large current densities and pressure gradients in the pedestal region. ELMs lead to a repeating loss of the plasma confined in the edge region, and particles and heat are lost to the wall on a time scale of $\lesssim 1\,$ms \cite{Zohm1996,Connor1998,Leonard2014}. During ELMs, filamentary eruptions are observed, as well as magnetic field stochastization \cite{Ham2020}. Magnetic perturbations during ELMs are linked to reconnection, since they are the result of resistive peeling-ballooning modes, which trigger magnetic reconnection at the resonant surfaces. 

To some degree, ELMS can be viewed as the analog of coronal emerging flux scenarios in laboratory plasmas, and it is worthwhile discussing briefly the commonalities and differences between laboratory and astrophysical plasma eruptions, on the example of ELMs in tokamaks and solar flares.

Both, solar corona and tokamak plasmas, are characterized by an electrical resistivity that is low in the sense that the Lundquist number is much larger than unity if the length scale considered is the macroscopic system size \cite{Fundamenski2007}. The macroscopic lengths are also much larger than the kinetic scales (electron and ion Larmor radii) in both systems. The strong toroidal field in tokamaks ensures that the plasma beta $\beta = p/(B^2/2\mu_0)$ (with $p$ the plasma pressure and $B$ the magnetic field) is smaller than unity, as it also holds in the flaring corona. Of course, there are enormous differences in absolute system size and duration of eruptive events, which both are larger by a factor of about $10^7$ in flares than in ELMs \cite{Fundamenski2007}. Yet, when considering the kinetic scales, which definitely are relevant for particle acceleration and heating, the respective plasma parameters have ratios of solar coronal values to tokamak values rather close to unity \cite{Fundamenski2007,McClements2019}.

There is an important topological difference between the two systems. In tokamak plasmas, the magnetic field lines in almost the entire plasma form a set of closed, nested, toroidal surfaces, whereas in flare plasmas, the coronal magnetic field is anchored in the much denser and cooler convection zone. As \citet{McClements2019} notes, there is always an outermost surface of closed magnetic flux in tokamaks, beyond which there is a relatively thin scrape-off layer (SOL) of plasma exhausted from the closed flux region. In most currently operating tokamaks, the magnetic field lines in the SOL are connected to a solid surface at the top or bottom of the vacuum vessel, called divertor. The SOL magnetic field topology thus resembles somehow to that of a flaring coronal loop, with the divertor playing the role of the convection zone. We though note that this is rather an analogy than a one-to-one correspondence, since the divertor is a passive device and does not drive the magnetic field in any way, whereas the convection zone is the main driver and ultimate energy source for coronal magnetic activity, such as flares.

A different, but still qualitative view, is to relate the solar convection zone with the well-confined region within the last closed magnetic surface of a tokamak, and to view the SOL as the analogue of the solar corona, into which ELMs break out from the well-confined region and cause eruptive events, very much like magnetic flux that emerges from the solar convection zone into the solar corona and leads to destabilization and eventually to explosive events, such as flares. In this view, turbulence in the well confined region of a tokamak and in the solar convection zone have in common to be the driver of the eruptive events, ELMs and flares, respectively.

\citet{Isliker2022} presented test-particle simulations of electrons during a nonlinear MHD simulation of an ELM. Their aim was to explore the effect of an eruptive plasma filament on the particle dynamics. They found that the electrons are moderately heated and accelerated during the ELM, on a fast time scale of the order of 0.5 ms. Also, the distribution of the kinetic energy exhibits a non-thermal tail, which is of power-law shape, reaching up to $90\,$keV. The acceleration exclusively takes place in the direction parallel to the magnetic field, and they showed that the parallel electric field is the sole cause of the particle acceleration. Most particles that escape from the system leave at one strike-line in the bottom region of the device (the outer divertor leg). The escaping high energy electrons in the tail of the energy distribution have characteristics of runaway electrons. The mean square displacement in energy space indicates that transport is super-diffusive, and, when viewing the acceleration process as a random walk, they found that the tails of the distributions of energy increments are of exponential shape. They also noted that transport in energy space is equally important of diffusive (stochastic) and convective (systematic) nature.

\begin{figure}[ht!]
	\centering
	\includegraphics[width=0.80\columnwidth]{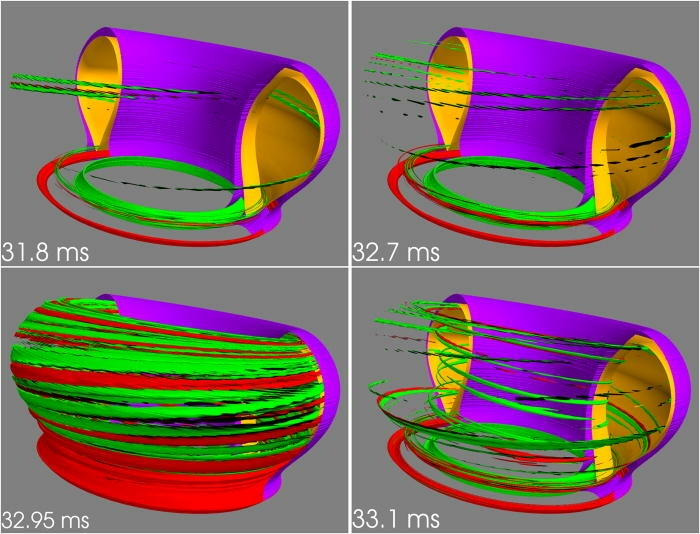}
	\caption{ Iso-contours of the parallel electric field $E_{||}$ (positive at $+10\,$V/m in red, and negative at $-10\,$V/m in green), for four different times  during the ELM, together with the separatrix (orange) and the plasma-boundary (violet) (both surfaces are half cut out). Reproduced with permission from Isliker  et al., Physics of Plasmas, {\bf 29}, 124 (2022), Copyright  2022 AIP.}
	\label{Heinz1}
\end{figure}

Isliker et al. also analyzed the MHD simulations per se (see Fig. \ref{Heinz1}), with the main finding that the histograms of the parallel electric field in the edge region adopt power-law shapes, and they concluded that this clearly non-Gaussian statistics is one of the main reasons for the moderately anomalous phenomena of particle transport in energy space that they find.

Solar wind \cite{Matthaeus15}, the Earth's magnetosheath \cite{Yordonova16,Voros17},  astrophysical jets \cite{Manolakou99, Comisso19, Meringolo23}, and edge localised turbulence in tokamaks \cite{Galassi19} are only a few space and laboratory examples of CoSs formation in strongly turbulent plasmas.

\section{Formation of Coherent structures from the fragmentation of large scale current sheets}\label{CoSCS}

The formation and fragmentation of large scale CSs in many space and astrophysical settings are worthwhile being studied in detail, as we will see at the end of this section.  
The fragmentation of an isolated large scale CS in 2D magnetic topologies, with the formation of plasmoids during the linear phase of the plasmoid instability, has been analysed extensively \cite{Comisso16, Comisso17}. Our main interest here is though in the 3D evolution and fragmentation of CSs in strongly turbulent environments.

Matthaeus and Laskin \cite{Matthaeus86} were the first to move away from the laminar reconnection flows and to explore the evolution of a 2D periodic CS in the presence of low level broadband fluctuations. It is worthwhile quoting the main findings from the abstract of their seminal article with the title {\it ``Turbulent magnetic reconnection"}: { \it``Nonlinear features of the evolution, appropriately described as turbulence, are seen early in the solutions and persist throughout the runs. Small scale, unsteady coherent electric current and vorticity structures develop in the reconnection zone, resulting in enhanced viscous and resistive dissipation. Unsteady and often spatially asymmetric fluid flow develops. Large scale magnetic islands, produced by reconnection activity, undergo internal pulsations. Small scale magnetic islands, or bubbles, develop near the reconnection zone, producing multiple X points. Large amplitude electric field fluctuations, often several times larger than the reconnection electric field, are produced by large island pulsations and by motion of magnetic bubbles. Spectral analysis of the fluctuations shows development of broad band excitations, reminiscent of inertial and dissipation range spectra in homogeneous turbulence. Two dimensional spectra indicate that the turbulence is broadband in both spatial directions. It is suggested that the turbulence that develops from the randomly perturbed sheet bears  a strong resemblance to homogeneous magnetohydrodynamic turbulence, and that analytical theories of reconnection must incorporate these effects."} 

\begin{figure}[h]
	\centering

	\includegraphics[width=0.70\columnwidth]{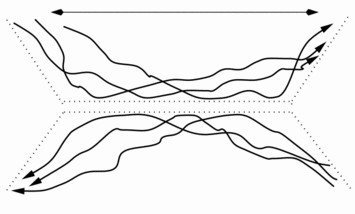}
	\caption{The  Lazarian and Vishniac  model \cite{Lazarian99} for a 3D weakly turbulent reconnection driver. The width of the outflow can be compatible with the large scale  characteristic of the turbulence due to stochastic wandering of field lines and the fragmentation of the initial current sheet. Reproduced with permission from Lazarian and Vishniac, The Astrophysical Journal, {\bf 517}, 700 (1999), Copyright  1999 AAS.} 
	\label{Lazarian99}
\end{figure}

Lazarian and Vishniac \cite{Lazarian99} returned to the formation of a CS in the middle of weakly turbulent flows (see their cartoon in Fig.\ \ref{Lazarian99}). They generalised the results of Matthaeus and Laskin \cite{Matthaeus86}, using a 3D magnetic topology, and stressed the importance of  stochastic magnetic field wandering, which causes fragmentation of  the initial large scale astrophysical current sheet. The reconnection in small fragments of the magnetic field determines the local reconnection rate and the local strength of the electric field.  The global reconnection rate is substantially larger, as many independent fragments reconnect simultaneously. Lazarian and Vishniac   also obtained  quantitative predictions for the reconnection rate (see details in \citet{Lazarian20}).

Onofri et al. \cite{Onofri04, Onofri06} numerically solved the incompressible, dissipative, magnetohydrodynamics
(MHD) equations in dimensionless units
in a three-dimensional Cartesian domain,
with kinetic and magnetic Reynolds numbers $R_v=5000$ and $R_M=5000$.
They set up the initial condition in such a way as to have a plasma that is at rest,
in the frame of reference of the computational domain, permeated by a
background magnetic field sheared along the $\hat{x}$ direction,
with a current sheet in the middle of the simulation domain.
They perturb these equilibrium fields with three-dimensional divergenceless large amplitude 
fluctuations. (For details about this MHD simulation see \citet{Onofri04}.)

\begin{figure}[h!]
\centering
\includegraphics[width=0.50\columnwidth]{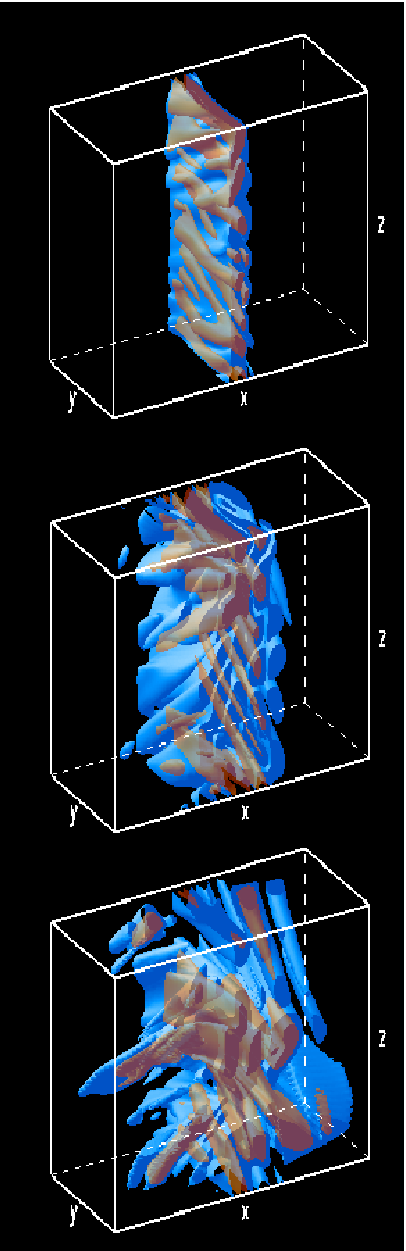}
\caption{Electric field iso-surfaces at $t=50\tau_A$,
$t=200\tau_A$ and $t=400\tau_A$. Reproduced with permission from Onofri  et al., Physical Review Letters, {\bf 96}, 151102 (2006), Copyright  2006 APS.}
\label{figure1}
\end{figure}

The nonlinear evolution of the system is characterized by the formation
 of small scale structures, especially on the lateral
regions of the computational domain,
and coalescence of current filaments in the center.
This behavior is reflected in
the three-dimensional structure of the electric field,
which shows that
the initial equilibrium is destroyed by the formation of current filaments.
After about $t=50 \tau_A$ (where $\tau_A$ is the Alfv\'en time),
the current sheet starts to be fragmented, as can be seen
in Fig.\ \ref{figure1}, where we show the configuration of the electric field
${\bf E}=\eta{\bf J}-{\bf v}\times{\bf B}$, calculated from the MHD simulation data.
The iso-surfaces of the electric field in Figure \ref{figure1} are shown for
different times, and they are calculated for two different threshold values of the electric field: the red surfaces represent higher values and the
blue surfaces represent lower values.
The structure of the electric field is characterized by small regions of space
where the  field is stronger, surrounded by larger volumes occupied
by lower electric field values.
At later times, the fragmentation is more evident,
and at $t=400\tau_A$,  the initial current sheet has been
completely destroyed and the
electric field is highly fragmented. To give a measure of the fragmentation of the electric field, Onofri et al. \citep{Onofri06} calculated the fractal dimensions of the fields shown in Fig. \ref{figure1}, using the box counting definition of fractal dimension. The applied thresholds are the same as those that have been used to draw the isosurfaces shown in Fig. \ref{figure1}. For the fields represented by the blue surfaces in Fig. \ref{figure1}, they found fractal dimensions $d=2$, $d=2.5$, and $d=2.7$ at $t=50\tau_A$, $t=200\tau_A$, and $t=400\tau_A$, respectively. For the more intense electric fields (red surfaces in Fig. \ref{figure1}), the fractal dimensions are $d=1.8$, $d=2$, and $d=2.4$ at $t=50\tau_A$, $t=200\tau_A$, and $t=400\tau_A$, respectively. These fractal dimensions can be considered as a way to quantify the degree of fragmentation of the electric field and to characterize the fraction of space that it fills as it evolves in time.

\begin{figure}[h!]
\centering
\includegraphics[width=0.70\columnwidth]{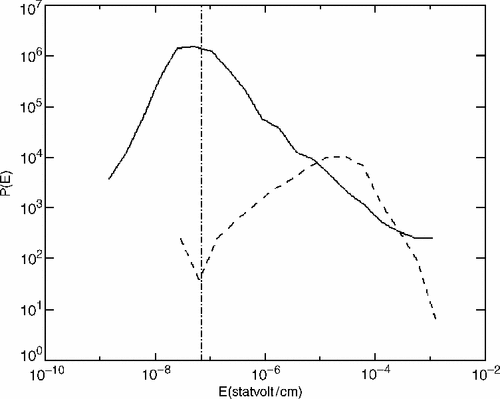}
\caption{Distribution function of the resistive (solid line) and convective (dashed line) electric field at $t=50 \tau_A$. The vertical line represents the value of the Dreicer field in the solar corona. Reproduced with permission from Onofri  et al., Physical Review Letters , {\bf 96}, 151102 (2006), Copyright  2006 APS.}
\label{OnofriE}
\end{figure}

Onofri et al \cite{Onofri06} calculate the magnitude $|\mb{E}|$ of the electric field at each gridpoint of the simulation domain and construct the distribution function of these quantities, which is shown in Fig.\ \ref{OnofriE} for $t=50\tau_A$. They separately plot the resistive and the convective component of the electric field. The resistive part is less intense than the convective part, but it is much more important in accelerating particles \cite{Onofri06}.

The fragmentation of a large scale CS was analysed by several authors \cite{Kowal11, Daughton11, Karimabadi13a, Oishi15, Dahlin15, Kowal17, Adhikari20}. Dahlin et al. \cite{Dahlin15}, using  kinetic simulations of 3D collisionless plasma with a guide field, analyze the fragmentation of a current sheet and the formation of small  scale filaments with strong electric fields. 
A different mechanism to reach the fragmentation of a large scale CS  is the presence of other CoSs in the surrounding of the CS, e.g.\ multiple reconnection sites \cite{Burgess16b}. 

%\begin{figure}[h!]
%\centering
%\includegraphics[width=0.30\columnwidth]{mcsa.eps}
%\includegraphics[width=0.30\columnwidth]{mcsb.eps}
%\caption{Evolution of multiple current sheets system at different %times.  (a) no guide field, (b) with guide field. The %configuration of the initial current sheets is shown in the left %panels with orange \cite{Burgess16}.}
%\label{figure2}
%\end{figure}

\begin{figure}[h!]
\centering
\includegraphics[width=0.80\columnwidth]{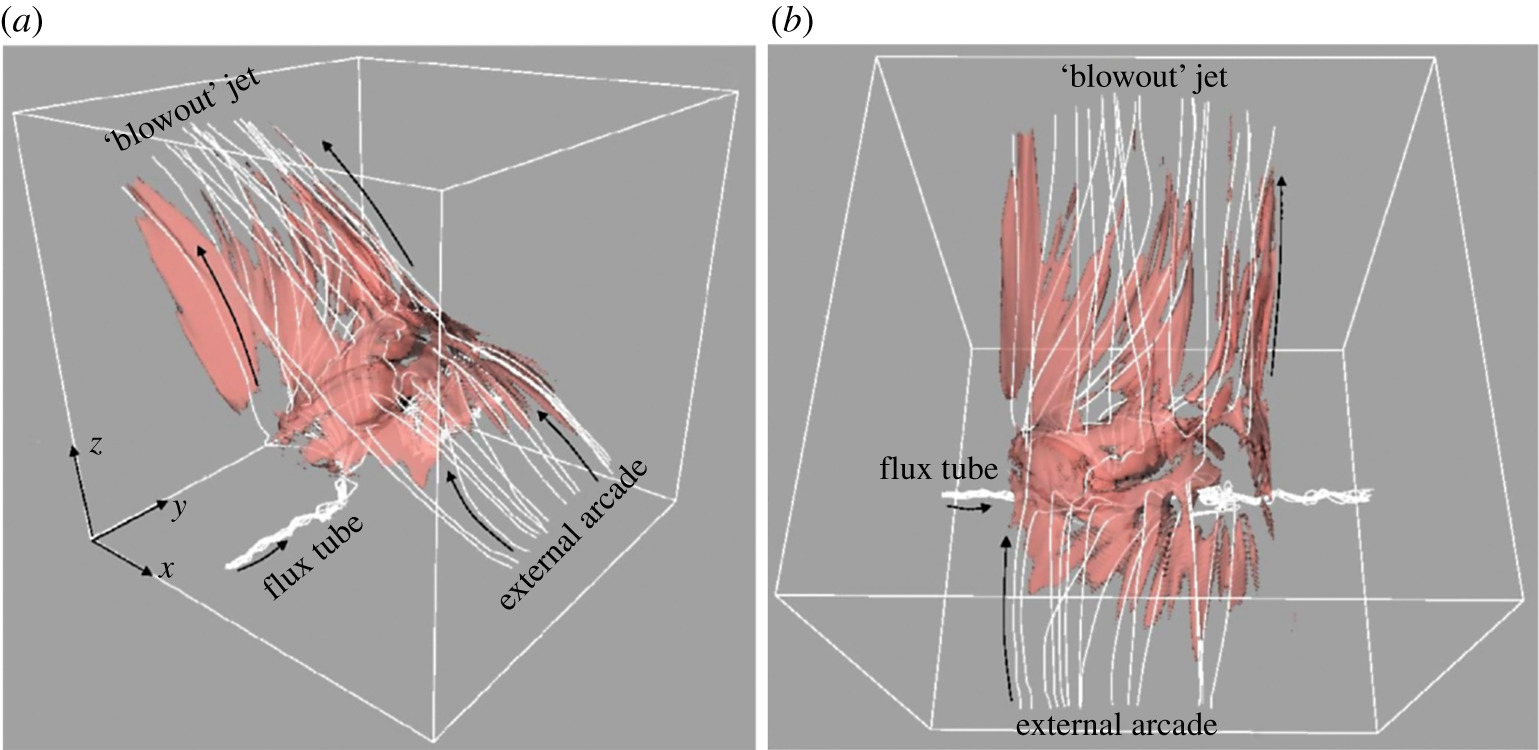}
\caption{Side-view (panel (a)) and top view (panel (b))  of the 3D field line topology and the velocity during the blowout jet emission. The direction of the field lines is shown by black arrows. Reproduced with permission from Archontis   et al., The Astrophysical Journal Letters , {\bf 769}, L21 (2013), Copyright  2006 AAS. }
\label{Archontis13}
\end{figure}

Greco et al. \cite{Greco16}, using Cluster high-resolution data, investigated the structure of thin current sheets that populate the turbulent solar wind. They concluded that in the solar wind, the turbulent cascade naturally forms current sheets at several scales, down to the proton skin depth scale. When approaching smaller scales, a current fragmentation process arises.

Beg et al. \cite{Beg22} analyzed the formation and evolution of a large-scale CS with a 3D MHD simulation of two merging flux ropes. They discovered that these systems exhibit self-generated and self-sustaining turbulent reconnection, which is fully 3D and fast.

Heyvaerts et al.\cite{Heyvaerts77} suggested that  magnetic flux emerging from below the photosphere and interacting with overlying magnetic fields, forms a quasi static large scale CS that may trigger a solar flare. Several numerical studies further explore the idea of the formation of a large scale CS from  emerging magnetic flux \cite{Archontis04,Archontis05,Galsgaard05, Archontis12a, Archontis12b, Moreno-Insertis13,Raouafi16, Wyper16, Wyper17}.

Archontis and Hood \cite{Archontis13} used a 3D resistive MHD code to follow the emergence of new magnetic flux into the corona with a pre-existing magnetic field. The formation and subsequent fragmentation of the large-scale reconnecting current sheets is obvious in their numerical study (see Fig. \ref{Archontis13}).

\begin{figure}[h]
\centering
\includegraphics[width=0.80\columnwidth]{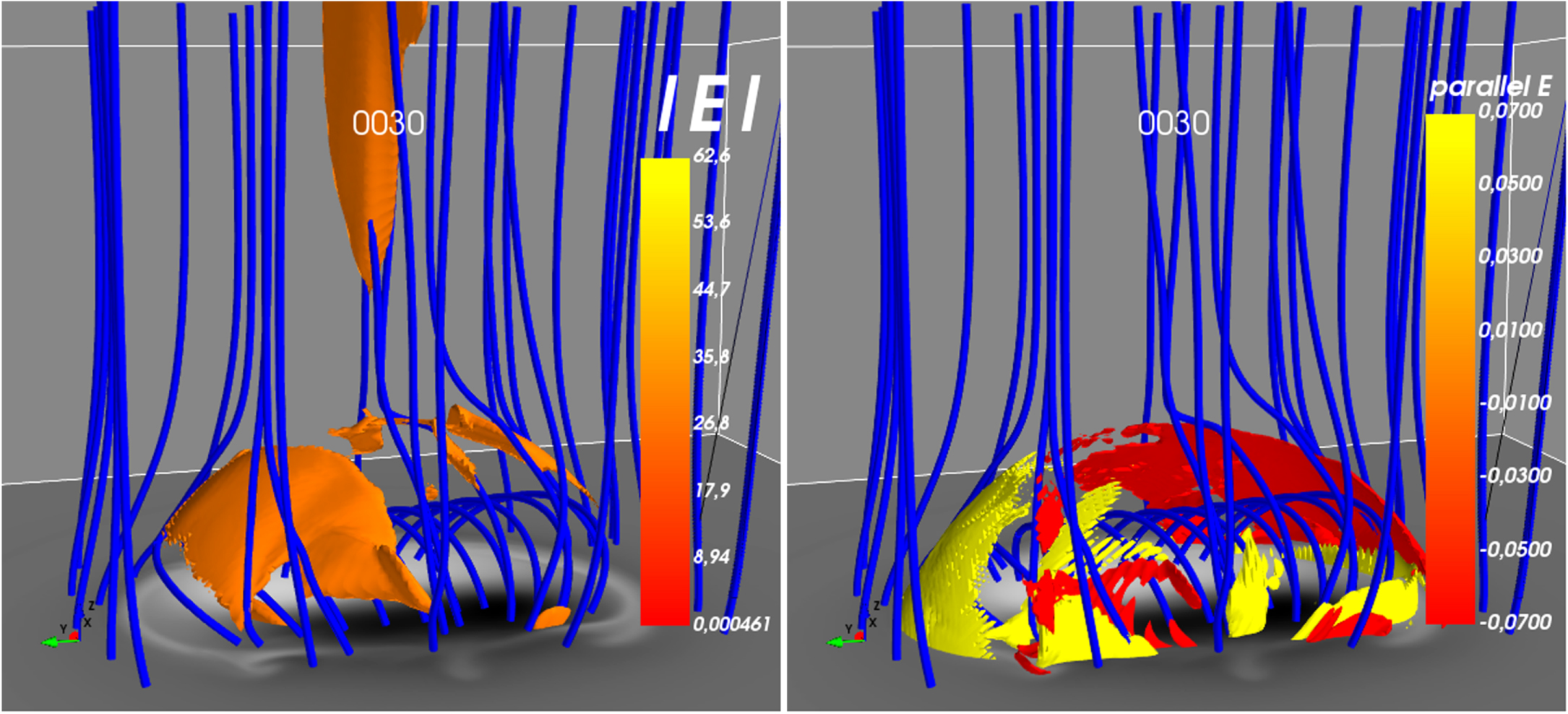}
\caption{Results from the MHD simulations: a close-up of the coronal part. The left panel shows a visualization of selected magnetic fieldlines (blue) together with an isocontour plot of the total electric field (orange 3D isosurfaces). The vertically oriented isosurface (orange) is aligned with the direction of the reconnected fieldlines and it indicates the emission of the standard jet. The $x$ $y$-plane at the bottom shows the photospheric component $B_z$ as a 2D filled contour plot. The electric field is in physical units [V m$^{-1}$]. In the right panel, isocontours of the parallel electric field are shown, indicating the fragmentation of the current sheet at the interface between the interacting magnetic fields. Reproduced with permission from Isliker   et al., The Astrophysical Journal, {\bf 882}, 57 (2019), Copyright  2019 AAS.} 
\label{Arch}
\end{figure}

Isliker et al. \cite{Isliker19} used the 3D MHD simulations of Archontis \& Hood \cite{Archontis13}, but focused on the statistical properties of the electric fields  in the vicinity of the fragmented large-scale current sheet. %(Fig. \ref{Arch}). 
The appearance of current fragmentation is apparent at the snapshots following the formation of the jet (see Fig.\ \ref{Arch}). The parallel electric field shows fragmented structures and has preferred regions of positive and negative sign. The fragmentation needs to be quantified with the use of cluster analysis and fractal dimension estimate.

\begin{figure}[h]
\centering
\includegraphics[width=0.60\columnwidth]{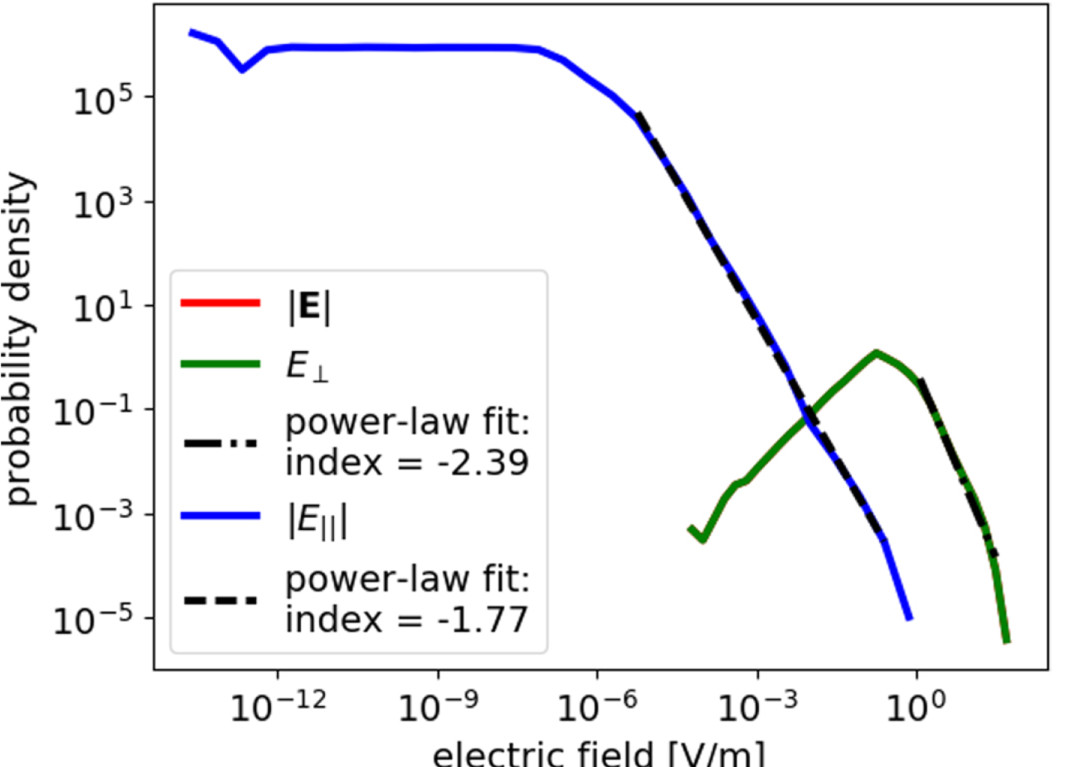}
\caption{MHD simulations, coronal part only, showing the distribution of the electric field from all coronal grid points, for the magnitude of the total electric field, the perpendicular component (they practically coincide), and the parallel component, respectively. The electric field is in units [V m$^{-1}$], and the mean Dreicer field is $4.6 \times 10^{-4}$ V m$^{-1}$.Reproduced with permission from Isliker   et al., The Astrophysical Journal, {\bf 882}, 57 (2019), Copyright  2019 AAS.}
\label{Archo2}
\end{figure}

Fig.\ \ref{Archo2} shows the histogram of the magnitude of the total electric field $| {\mathbf{E}}| $, the parallel $| {E}_{||}| $, and the perpendicular $E_\perp$ component of the electric field, determined from all coronal grid points. They all show a power-law tail with a rollover at high values. The power-law index of the fit is -1.8 for the parallel electric field, and -2.4 for the total and the perpendicular electric field,  just that the total and perpendicular electric field attain larger values. In any case, the parallel electric field is two orders of magnitude smaller than the total electric field, which thus basically coincides with the perpendicular electric field. Also, the parallel electric field shows a much more extended power-law tail than the perpendicular and the total one. Isliker et al. \cite{Isliker19} conclude that power-law-shaped distributions are inherent to the electric field and its components. Similar results have also been found in MHD simulations of a decaying current sheet, as we reported  earlier in this section (see Fig.\ \ref{OnofriE}).

%Histograms of the total, parallel and perpendicular electric %field suggest that the perpendicular electric filed is super %Dreicer but the parallel electric field, only a fraction of the %grid points (see Fig. \ref{Arch2} ), this sets the threshold for %the iso-contours, clusters and fractal dimension analysis. 
%\begin{figure}[h]
%\centering
%\includegraphics[width=0.90\columnwidth]{arch5.eps}
%\includegraphics[width=0.45\columnwidth]{arch4.eps}

%\includegraphics[width=0.45\columnwidth]{Dtr1.eps}
%\includegraphics[width=0.45\columnwidth]{Dtr2.eps}
%\caption{(a) Isocontures of the parallel electric field. (b) %Cluster analysis}
%\label{Arch2}
%\end{figure}

\begin{figure}[h]
\centering
\includegraphics[width=0.80\columnwidth]{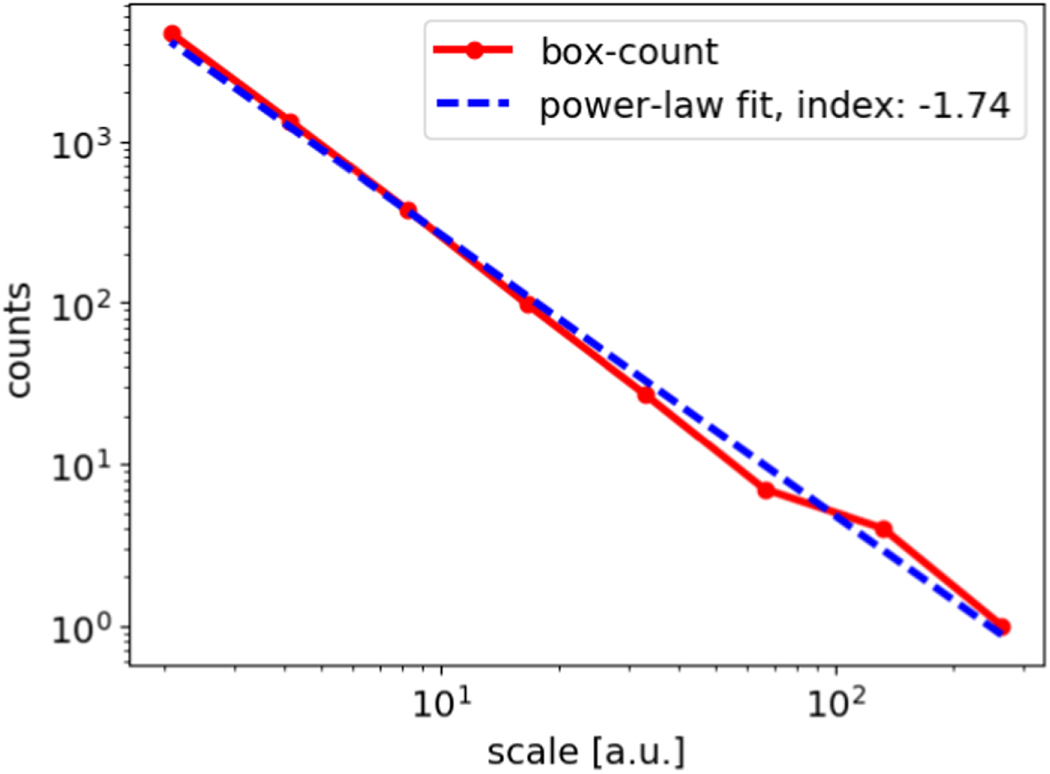}
\caption{Fractal dimension of the parallel electric field: scaling of the 3D box-counting algorithm. Reproduced with permission from Isliker   et al., The Astrophysical Journal, {\bf 882}, 57 (2019), Copyright  2019 AAS.}
\label{Archo3}
\end{figure}

Isliker et al. \cite{Isliker19} also investigate  the spatial structure of the parallel electric field, applying cluster analysis and calculating its fractal dimension. They consider the magnitude of the parallel electric field $| {E}_{\parallel }| $ at all coronal grid points, and apply a threshold below which $| {E}_{\parallel }| $ is set to zero. For the threshold, they use the same value of 0.07 as for the isocontours of $E_\parallel$ in Fig. \ref{Arch}.
They define a cluster as a set of grid points with (a) an above-threshold value of $| {E}_{\parallel }| $ at all the grid points belonging to the cluster, and (b) the cluster's grid points are connected through their nearest neighborhoods in 3D Cartesian coordinates. It follows that a cluster is surrounded by grid points with below-threshold $| {E}_{\parallel }| $. They found that there are 162 clusters, and 2 of them are very dominant in spatial extent, one corresponding to the positive and one to the negative extended parallel electric field region in Fig. \ref{Arch}.
Using the same data as employed in the cluster analysis (the magnitude of the parallel electric field $| {E}_{\parallel }| $ at all the coronal grid points, set to zero when below the threshold value of 0.07), Isliker et al. \cite{Isliker19} apply a standard 3D box-counting method in order to determine the fractal dimension $D_F$ of the region with above-threshold parallel electric field. Fig. \ref{Archo3} shows the scaling of the box counts with the box scale, where there is a clear power-law scaling in the entire range, whose index, per definition of the box-counting method, equals the fractal dimension, so they find $D_F = 1.7$ for snapshot 30 (standard jet) and $D_F = 1.9$ for snapshot 53 (blowout jet).

The regions of high parallel electric field can thus be interpreted as thinned out 2D sheets, as it also corresponds to the visual impression that is given by Fig. \ref{Arch}. Also, the "filling-factor" (fractal dimension) is higher at the blowout jet compared to the time when the standard jet is emitted.  After all, the spatial structure of the regions of strong parallel electric field can be characterized as fragmented and fractal, with the various cluster-size distributions exhibiting double power-law scalings.

The formation of large scale 3D CSs in the middle of converging strongly turbulent flows is present in several space and astrophysical phenomena, e.g.\ the (turbulent) magnetotail \cite{ElAlaoui21} driven by the turbulent solar wind, the (turbulent) magnetopause \cite{ChenLJ21} driven by the turbulent magnetosheath,  eruptive solar flares (resulting from emerging magnetic flux \cite{Archontis13,Nishida13}, or from the eruption of large scale magnetic topologies \cite{Inoue16, Inoue18, Cheng18}), and the outer helioshere \cite{Burgess16b}. In all these cases, the fragmentation of the CS and its replacement with a much larger strongly turbulent region has led  observers to report as a CS the relatively large scale 3D region where the CS's fragmentation was initiated by CoSs included in the 3D magnetic topology of   randomly wandering magnetic field lines\cite{Lazarian99}.  

\section{Formation of Coherent structures in the vicinity of large scale Shocks}\label{CoSShocks}

\begin{figure}[h!]
\centering
\includegraphics[width=0.60\columnwidth]{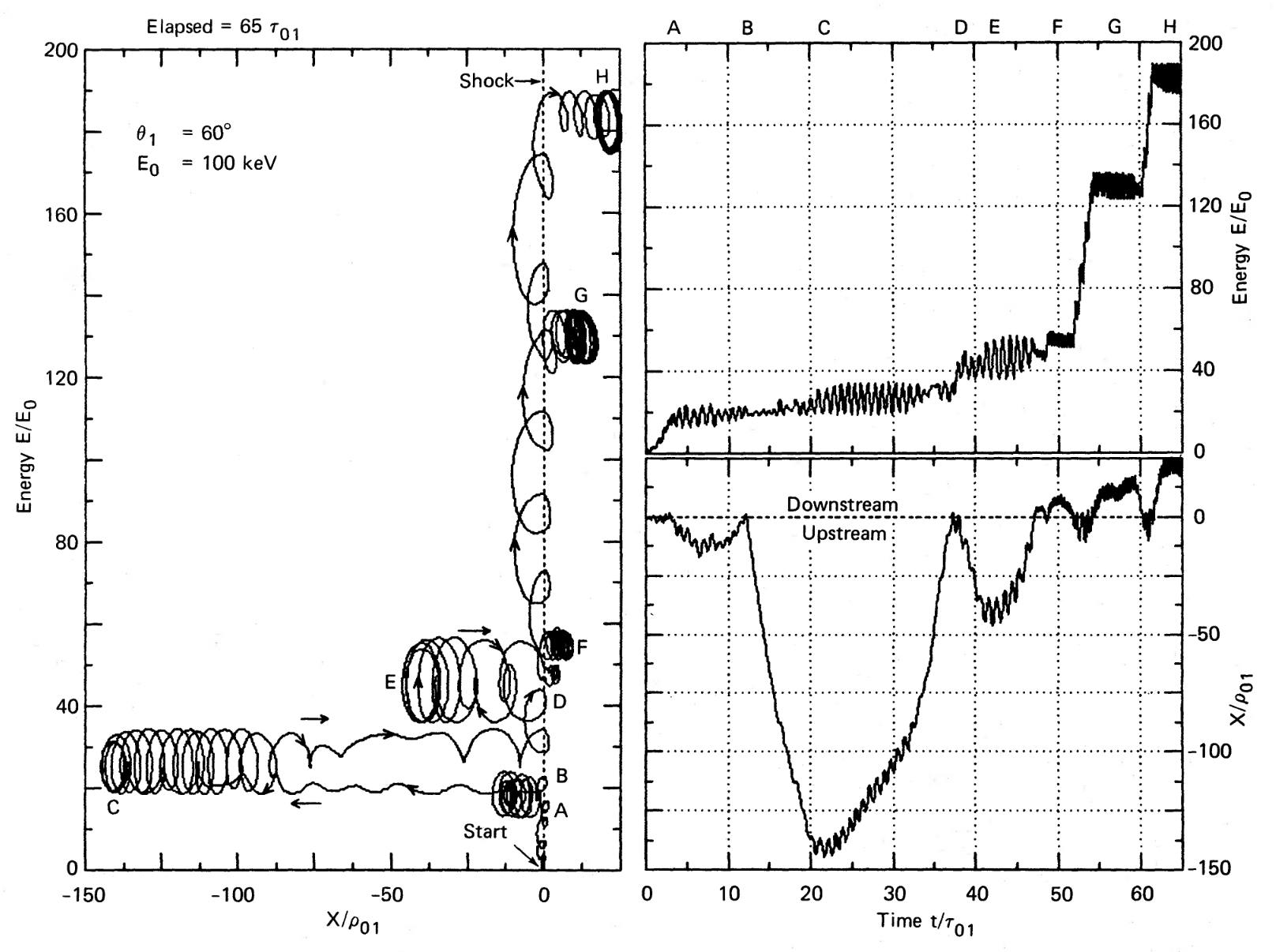}\caption{Sample orbit for a quasi-perpendicular shock ($\theta= 60^o$). Left: Energy vs.\ $X$  in the shock frame, where $X$ is the distance from the shock ($X < 0$ upstream; $X > 0$ downstream).  Right: Evolution of the energy (upper panel) and of the $X$-component of the particles position (lower panel) as a function of time. See details in the article \citet{DeckerVlahos86}. Reproduced with permission from Decker and Vlahos, The Astrophysical Journal, {\bf 306}, 710 (1986), Copyright  1986 AAS.}
\label{DeckerVlahos86}
\end{figure}

The theoretical analysis of large scale 2D quasi-static shock waves followed, for many years, the same steps as deployed in the large scale Sweet-Parker model for CSs \cite{Parker57}. A shock discontinuity was placed by hand in the middle a the simulation box, and weak turbulence was added upstream and downstream, to play the role of converging passive scatterers, since the energization of particles is due to their crossing of the shock discontinuity\cite{Decker88} (see Fig. \ref{DeckerVlahos86}). The formation of the shock and the way the specific normal modes were exited were not discussed in the initial model proposed by Fermi in 1954\cite{Fermi54}, whose approach was used as the base for the subsequent theoretical studies \cite{Drury83}. In this  simplistic scenario, the low amplitude wave modes are hand picked carefully to be able to scatter efficiently the particles. The particles, in order to interact with the waves upstream and downstream,  should have initial velocity higher than the phase velocity of the wave. This is the well known injection problem, for which a pre-acceleration mechanism is necessary \cite{Drury83}. 

Finally, weak turbulence upstream of a shock cannot confine efficiently the accelerated particles in the vicinity of the shock. It was then proposed that streaming  instabilities, driven by energized particles upstream  of a shock,  can be the solution for the trapping of particles in the vicinity of a shock \cite{Bell78a, Bell78b}.

Obviously, the formation and evolution of large scale shock waves in space and astrophysical explosions  encounter pre-existing or self-generated strong turbulence (e.g.\ the Earth's and planetary bow shocks in the solar wind, or the shocks formed by Coronal Mass ejection in the heliosphere, or the Super Nova (SN) explosions in astrophysical plasmas). The strongly turbulent plasma upstream  of a large scale shock  carries a variety of CoSs, which interact with the shock and play a crucial role in its evolution \cite{Burgess16a}.  When CoSs cross a shock discontinuity, they are amplified dramatically (see \cite{Omidi07, Karimabadi2014,  Gingell20, Lu20a, Lu20b,ChenLJ21, Kropotina21, Trotta21, Parks21} for the magnetosheath, and the references therein), and the shock surface shows large scale ripples \cite{Gingell17}.

\begin{figure}[h!]
\centering
\includegraphics[width=0.70\columnwidth]{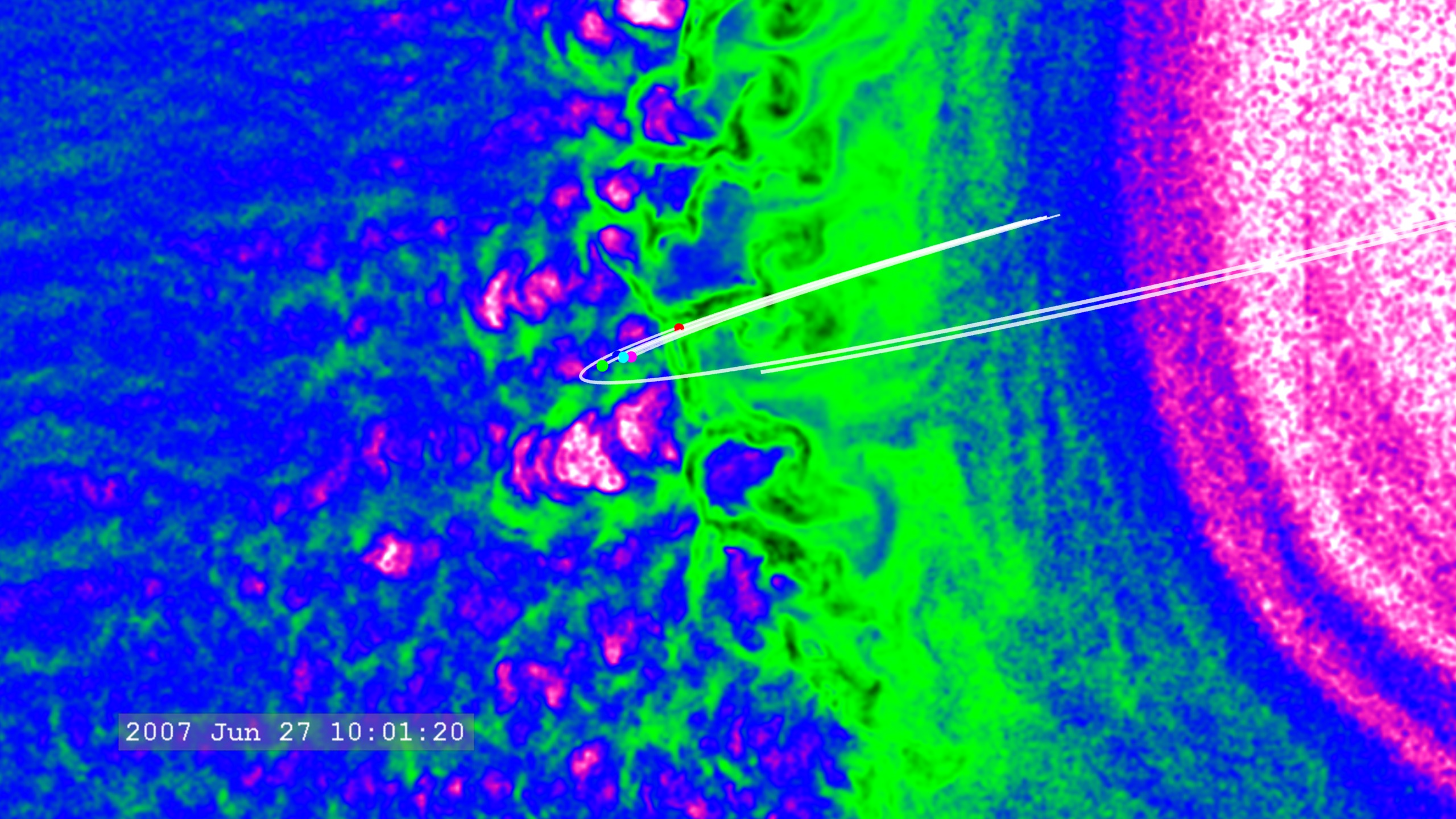}\caption{The five THEMIS satellites  moving along their orbits.  The 2-D data from the Omidi simulation \cite{Omidi07} are  faded in and a zoom is made in to a view of the satellites in the turbulent region near the bow shock. The 'cavitons' of violet and white color  illustrate a broad range of CoSs in this turbulent fore-shock region.Reproduced with permission from Omidi, AIP Conference Series, {\bf 932}, 181 (2007), Copyright  2007 AIP. }
\label{slams}
\end{figure}

The Earth's bow shock is the most studied environment of a collisionless large scale shock due to the availability of {\it in situ} data. The turbulent solar wind upstream carries several types of  CoSs (CSs, Cavitons, Rotational Discontinues (RDs), Short Large Amplitude Magnetic Structures (SLAMs), etc., see Fig.\ \ref{slams}), and the instabilities excited by the solar wind ions, reflected at the shock front, re-inforce the unstable CoSs before they are convected downstream \cite{Burgers05, Omidi07, Wilson13, Palmroth18, Gingell20, Lu20a, Lu20b, Kropotina21}. 

Karimabadi et al. \cite{Karimabadi2014} use a global hybrid simulation (electrons as fluid, kinetic ions) to explore the formation of CoSs in the vicinity of the bow shock. They analyze the link of  large scale shock waves, turbulence upstream and downstream, and CoSs. The solar wind turbulence  upstream can easily reach the strongly turbulent level ($\delta B/B \approx 1$) and drive turbulent reconnection and the formation of large scale low frequency electromagnetic waves that are compressed and amplified as they cross the shock.

\begin{figure}[h!]
%\centering
%\includegraphics[width=0.60\columnwidth]{shock.png}\\
%\includegraphics[width=0.45\columnwidth]{Matsorbit.png}
%\sidesubfloat[]
{\includegraphics[width=0.7\columnwidth]{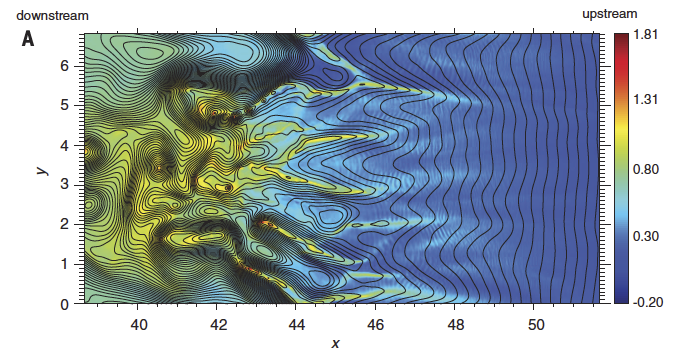}%
		\label{Shock1}}\hfill\\
		%\sidesubfloat[]
		{\includegraphics[width=0.6\columnwidth]{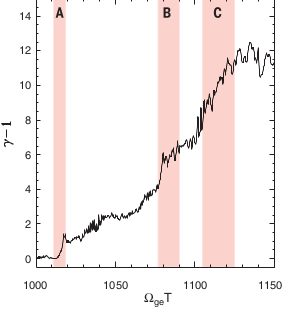}%
		\label{Shock2}}\hfill\\
		%\sidesubfloat[]{\includegraphics[width=0.6\columnwidth]{3DCS.jpeg}%
		%\label{MHD_t}}\hfill\\
\caption{(a) Supercomputer simulation of a strong collisionless shock, revealing the presence of CoSs.  (b)  An electron's kinetic energy $(\gamma-1)$ as
a function of time (normalized to the upstream electron gyro frequency).  The
times from (A) to (C), for which the particle's orbit is marked,  indicate its interaction with specific CoSs downstream (see details in Matsumoto et al. \cite{Matsumoto15}). Reproduced with permission from Matsumoto   et al., Sience, {\bf 347}, 974 (2015), Copyright  2015 AAAS. }
\label{figure3}
\end{figure}

Matsumoto et al. \cite{Matsumoto15} presented a supercomputer particle in cell (PIC) simulation, showing that strong collisionless shocks drive CoSs in the transition region (see Fig.\ \ref{figure3}a). They use high computational capacity  to follow the evolution of a collisionless  shock moving with Mach number $M_A \approx 40$. In Fig.\ \ref{figure3}a, between the upstream region $x>59$ and the downstream region $x<39$, a transition region is characterized by tangled magnetic field lines. This region is in the strongly turbulent state, and the shock front cannot be visually identified. The fragmentation of the shock front drastically modified the assumptions that are behind the analytical work and the simplistic models for diffusive shock acceleration \cite{Caprioli19, Caprioli20}.

It is also of interest to note that filamentary structures are
created, as can be seen from the density profile in
the transition region in Fig.\ \ref{figure3}a. These filaments are associated
with folded magnetic field lines, and the enhanced density
regions contain CoSs.  Matsumoto et al. \cite{Matsumoto15} also  observe
that magnetic reconnection takes place at
multiple sites, and, as a result, magnetic
filaments are formed along the initial current sheet.  Fig.\ \ref{figure3}b presents the evolution of the kinetic energy of a particle, which shows enhanced gains in energy during the interactions with specific CoSs downstream.
Similar evidence for magnetic reconnection is
found in other current sheets in the transition
and the downstream region \cite{Matsumoto15}. This fragmentation of the initial large scale shock front is reminiscent of the large scale CS fragmentation discussed in Sec.\ \ref{CoSCS}. Similar results are reported by Caprioli and Spitkovsky \cite{Caprioli14a} by using 3D hybrid simulations.

Korpotina et al. \cite{Kropotina21} analysed the interaction of CoSs (in their case they focused on  rotational discontinuities) carried by the turbulent solar wind with the Earth's Bow shock. They used {\it in situ} multispacecraft  observations and performed hybrid kinetic simulations. In their article, they stress the amplification of the CoSs in the vicinity of the shock discontinuity. The amplification of the   CoSs carried by the solar wind may be as high as two orders of magnitude. The amplification 
%of the CoSs carried by the solar wind 
in the foreshock may be due to  streaming instabilities, driven by the interaction of the solar wind with reflected solar wind particles. The Earth's bow shock crossing affects the CoSs, since downstream of the shock the magnetic field and the plasma density are amplified (actually, the observed compression ratio far exceeds the Hugoniot prediction; see also the recent study by Trotta et al \cite{Trotta22}). Guo et al. \cite{GuoZ21}, using a 3D global hybrid simulation, showed that the upstream turbulence at a quasi-parallel shock may intensify the presence of CoSs at the rate of reconnection at the  magnetosheath downstream of the Earth's bow shock (similar results are reported in several recent articles \cite{Lu20a, Lu20b, Gingell20, ChenLJ21, Trotta21,  Parks21}). 

Unfortunately, the multi-scale character and the complexity of the microphysics \cite{Burgess16a} present in an evolving large scale shock cannot be explored by current numerical simulations, and neither  can they follow a shock's evolution  for long time. This is the main reason why we are still missing many important details on the evolution of CoSs and their role in the heating and acceleration of particles.

The topic of the interaction of large scale shocks with plasma turbulence and the formation of CoSs upstream and downstream as well as the non-stationary evolution of a 3D shock discontinuity \cite{Gingell17} is currently an open problem in space and astrophysical plasmas.   Moreover, the collective interaction of particles with  the rich  variety of CoSs (and not only the reconnecting CSs) and its effect on the heating and acceleration of particles also remain an open problem. 

%\section{Statistical properties of coherent structures in strong %turbulent plasma}

%\part{Turbulent reconnection  inside the Solar Corona}
\section{Formation of coherent structures  in the solar atmosphere}\label{CoS_Shocks}

In the solar atmosphere, CoSs are formed through the strong  magnetic coupling with the turbulence in the  convection zone.  We can split our presentation of the formation of CoSs in the solar atmosphere in two parts: (1) The formation of CoSs by the random shuffling of the footpoints of the slowly changing  magnetic fields, and (2) the formation of CoSs  during the emerging magnetic flux and/or the large scale magnetic eruptions. 

\subsubsection{Formation of CoSs in quasi-static closed magnetic topologies}

\begin{figure}[h]
\centering
\includegraphics[width=0.70\columnwidth]{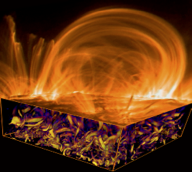}
\caption{The magnetic link of the convection zone with the solar corona: The turbulent convection zone drives the formation of CoSs in the closed magnetic topologies above the the photosphere.}
\label{ar2}
\end{figure}

In the solar  atmosphere, the formation of  CoSs (in the cited literature the emphasis is on the formation and disruption of current sheets) through small scale random shuffling of the footpoints of emerged magnetic fields, caused by the convection zone, was first proposed  by Gold \cite{GoldT64} in the 60s as the main mechanism for coronal heating. The initial conjecture  was developed further in the 70s and the 80s \cite{Parker72,Glencross75, Levine74, Sturrock81, Parker83}, and it continues till today, as we will show in this section. Based on this scenario, Parker\cite{Parker88} introduced the   concept of nanoflares for the formation and disruption of small scale current sheets  (Glencross \cite{Glencross75}   defined the same phenomenon as ``flarelike'' brightenings, and several years later Chiuderi \cite{Chiuderi93} called the same phenomenon  ''elementary events'' ) as   the main mechanism for coronal heating. In other words, the formation of small scale CSs and their collective Ohmic dissipation in closed magnetic topologies is widely accepted as a potential mechanism for coronal heating today.  In Fig.\ \ref{ar2}, we present a cartoon showing the magnetic coupling  of the turbulent convection zone with  the solar atmosphere above an active region. The details of the formation of the CoSs through the interaction of the solar atmosphere with the turbulent convection zone remain an open problem till today.  

\begin{figure}[h]
%\centering
% Use the relevant command for your %figure-insertion program
%\includegraphics[width=0.40\columnwidth]{FlareT%ime.png}
%\includegraphics[width=0.40\columnwidth]{fs.eps%}
%\sidesubfloat[]
{\includegraphics[width=0.5\columnwidth]{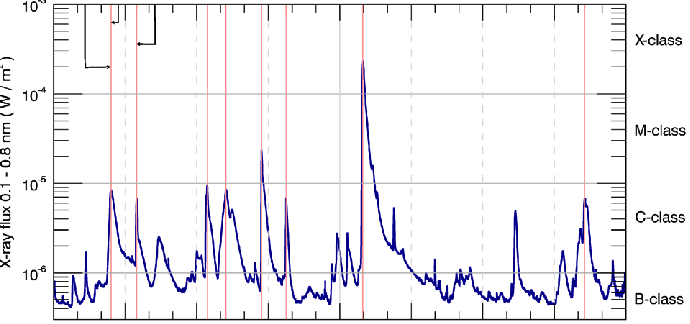}%
		\label{flaretime}}\hfill
		%\sidesubfloat[]
		{\includegraphics[width=0.5\columnwidth]{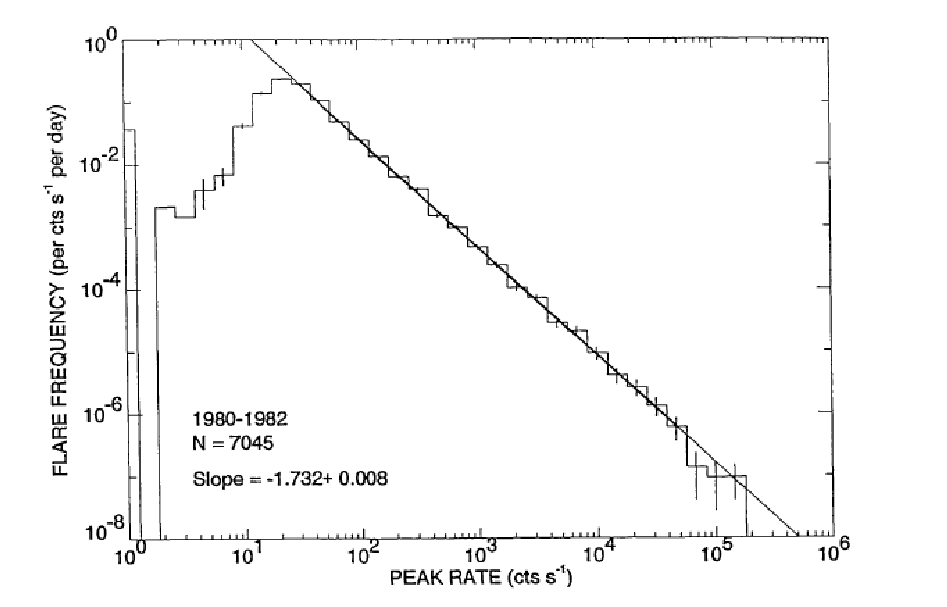}%
		\label{flarestat}}\hfill\\
		%\sidesubfloat[]{\includegraphics[width=0.6\columnwidth]{3DCS.jpeg}%
		%\label{MHD_t}}\hfill\\
%\vspace{5cm}       % Give the correct figure height in cm
\caption{(a) A typical solar X ray flux from February  14, 2011 till February 15, 2011. (b) The frequency distribution of the peak count rate for the flares recorded in 1980-1982. Reproduced with permission from Grosby, Non-linear Processes in Geophysics, {\bf  18}, 791 (2011), Copyright 2011, Licensed under Creative Commons Attribute (CC-BY).}
\label{fig:2}       
\end{figure}

In the beginning of the nineties, it was shown  that the peak-luminosity distribution of flares displays a
well-defined, extended power law with index \heinz{$-1.7$ to $-1.85$}, see Fig.\ \ref{fig:2}
\cite{Crosby93}. This analysis was repeated and confirmed by several authors over the last thirty years (see the reviews \cite{Crosby11, Aschwanden16}).

Deviations from the power law behavior appear at the lowest energies. It has been
pointed out that these deviations  are due to instrumental limitations. The interesting part from the statistical analysis of solar flares is that the power-law is extended to all flare energies (from micro flares  with energies around $10^{24}$  erg to flares with energies $10^{30}$ erg).  The magnetic link  of the turbulent driver (convection zone) with the solar atmosphere is probably behind the  statistical properties of the energy release in closed magnetic topologies and can serve as a test for the numerical simulations that will be presented in this section.

In the 90's,  the conjecture reported above for the formation of CSs in solar active regions, through random shuffling of the footpoints of closed magnetic fields,  was tested by using 2D and 3D numerical  MHD codes  \cite{Mikic89,  Einaudi96,  Galsgaard96, Galsgaard97a, Gasgaard97b, Georgoulis98, Rappazzo10,  Hansteen14, Dahlburg16, Kanella17, Kanella18, Hansteen19, Einaudi21}. 

Einaudi et al. \cite{Einaudi96} analyzed a 2D section of a coronal loop, subject to random forcing of the magnetic fields. The title of their article was ``Energy release in a turbulent corona'', and they discussed the spontaneous formation of current sheets.  Geourgoulis et al. \cite{Georgoulis98} extended the simulation of Einaudi et al.\ for much longer times, since their aim was to extract reliable statistical information on turbulent reconnection in the solar atmosphere. Their main result was that the distribution functions of both the maximum and average current dissipation, for the total energy content, the peak activity and the duration of such events are all shown to display robust scaling laws.  
This result was recovered by the analysis performed twenty years later by \citet{Zhdankin13}, which is briefly reported in section II. 

\begin{figure}[h]
\centering
\includegraphics[width=0.70\columnwidth]{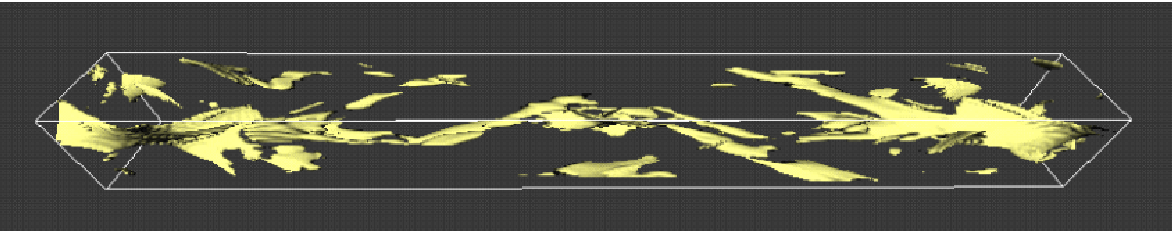}\\
\includegraphics[width=0.450\columnwidth]{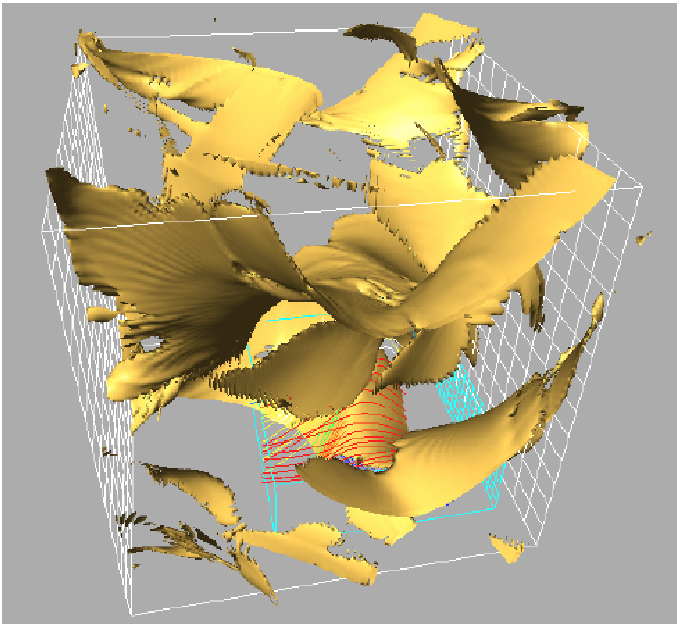}
\caption{(Top) Isosurfaces of Joule dissipation regions inside a solar magnetic loop driven by random shuffling of the footpoints. (Bottom)  Isosurfaces of strong electric currents in a snapshot. Reproduced with permission from Galsgaard and Nordlund, Journal of Geophysical Research, , {\bf 101}, 13445 (1996), Copyright  1996 Wiley.}
\label{Dtr}
\end{figure}

Galsgaard and Nordlund \cite{Galsgaard96, Galsgaard97a, Gasgaard97b} solved the dissipative 3D MHD equations in order to investigate the formation of CSs inside an  initially homogeneous magnetic flux tube, stressed by large scale sheared random motions at the two boundaries. The spontaneous formation of CSs at random places at random times inside the structure is shown in Fig. \ref{Dtr}. CSs of all scales appear and evolve since the large scale CSs fragment, as we have shown in section III.

Reconnection of the current sheet(s) will straighten the field lines, but will also cause  disturbances in the surrounding plasma, causing further fragmentation of the energy release processes. A strongly turbulent environment is established inside the magnetic flux tube, driven by the random shear caused by photosheric motions and by the energy delivered into the solar corona through the dissipation of  fragmented current sheets (see similar results in  \cite{Onofri06, Dahlin15}). The evolution of the CoSs  depends on the velocities of the boundary motions and the initial magnetic field strength in the magnetic flux tube.

Turkmani et al. \cite{Turkmani05, Turkmani06} analyzed the statistical properties of the electric fields inside the flux tube studied by Galsgaard and Nordlund \cite{Galsgaard96, Galsgaard97a}. The magnetic and electric fields are obtained from a 3-D MHD experiment that represents a coronal loop with photospheric regions at both footpoints. Photospheric footpoint motion leads to the formation of a hierarchy of  current sheets, see Fig.\ \ref{Turk}a. The distribution function of the resistive electric field has a power law tail for the super-Dreicer electric fields,  with  slope $-2.8$ (see Fig.\ \ref{Turk}b), and this finding is analogous with the results reported in other studies related with the fragmentation of a large scale CS \cite{Onofri06, Isliker19}.

\begin{figure}[ht!]
%\sidesubfloat[]
\includegraphics[width=0.6\columnwidth]{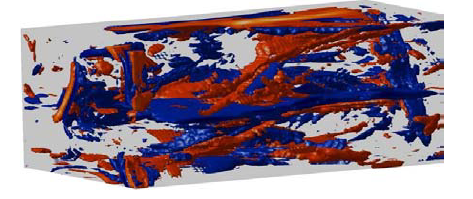}%
		%\label{Turk}}
	\hfill\\
		%\sidesubfloat[]
		{\includegraphics[width=0.6\columnwidth]{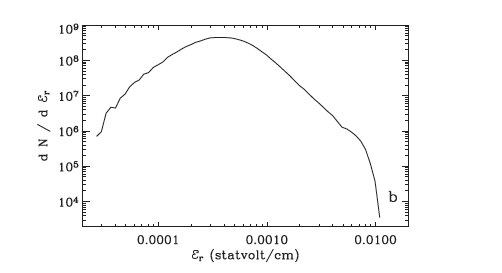}%
		\label{Turk}}\hfill
		%\sidesubfloat[]{\includegraphics[width=0.55%\columnwidth]{f1c}%
		%\label{MHD_t}}\hfill\\
	\caption{ (a) Snapshot of the resistive electric field configuration within the coronal volume, as calculated from the global MHD model. The blue and red regions represent electric field regions that point towards the left and right foot point, respectively.    (b) The distribution function of the resistive electric field. Reproduced with permission from Turkmani   et al., Astronomy and Astrophysics, {\bf 449}, 749 (2006), Copyright  2006 ESO.
			}\label{Turk}
\end{figure}

%\begin{figure}[h]
%\centering{}
%\includegraphics[width=0.55\columnwidth]{f22a.png} %\\ %
	%	%\label{turk1}}\hfill
		%\includegraphics[width=0.55\columnwidth]%{f22b.png}%
		%\label{turk2}}\hfill
		%\sidesubfloat[]{\includegraphics[width=0.6\columnwidth]{3DCS.jpeg}%
		%\label{MHD_t}}\hfill\\
%\centering
%\includegraphics[width=0.40\columnwidth]{Turk1.png}
%\includegraphics[width=0.40\columnwidth]{Turk2.png}
%\includegraphics[width=0.45\columnwidth]{Dtr1.eps}
%\includegraphics[width=0.45\columnwidth]{Dtr2.eps}
%\caption{  (a) Snapshot of the resistive electric %field configuration within the coronal volume, as %calculated from the global MHD model. The blue and %red regions represent electric field regions that %point towards the left and right foot point, %respectively.  (b) The distribution function of the %resistive electric field. \cite{Turkmani06}
%}
%\label{Turk}
%\end{figure}

\begin{figure}[h]
\centering
\includegraphics[width=0.80\columnwidth]{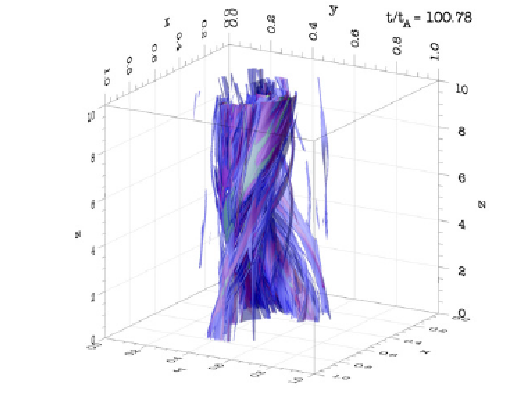}
\caption{Isosurfaces of the squared current $j^2$ at a time after the start of the kink instability. Reproduced with permission from Rappazzo et al., The Astrophysical Journal, {\bf 771}, 76  (2013), Copyright  2013 AAS. }
\label{Kink}
\end{figure}

In a series of articles, Rappazzo et. al. \cite{Rappazzo10, Rappazzo13, RappazzoParker13, Rappazzo17}, following the steps of the work of Galsgaard and Nordlund \cite{Galsgaard96}, analyzed several aspects of the formation  of CSs in the solar corona (see Fig.\ \ref{Kink}). The observational expectations of intermittent energy dissipation in the formed and fragmented CSs were also analyzed in depth \cite{Dahlburg16, Einaudi21}.

The numerical simulations presented so-far were performed, for the sake of simplicity, in isolated magnetic flux tubes. 

\begin{figure}[h!]
\centering
\includegraphics[width=0.60\columnwidth]{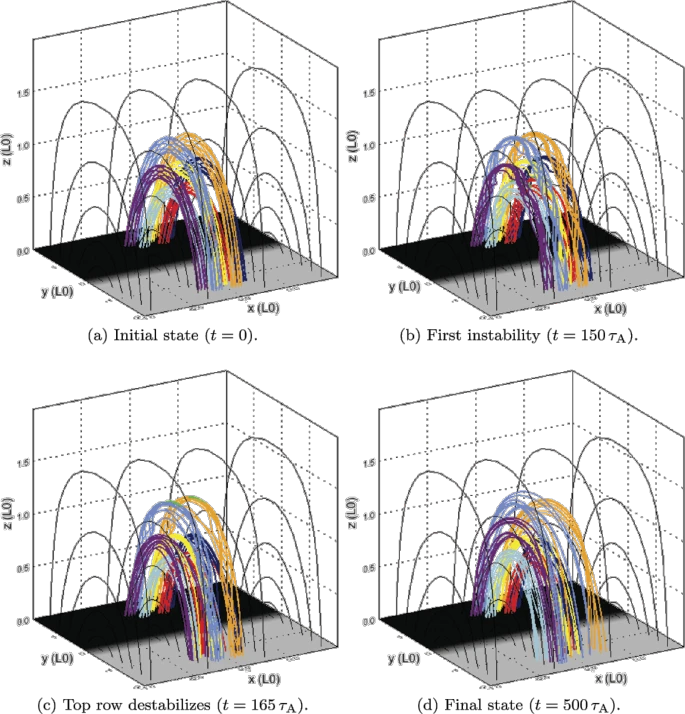}\\
\caption{ Seven threads with identical drivers: 3D magnetic configuration at different times. When one of the threads  is destabilised, it interacts with the rest of the threads and destabilizes all of them, causing the formation of several CoSs.  Reproduced with permission from Threlfal   et al., Solar Physics, {\bf 296}, 120 (2021), Copyright  2021 Springer}
\label{Thre}
\end{figure}

\begin{figure}[h!]
\centering
\includegraphics[width=0.90\columnwidth]{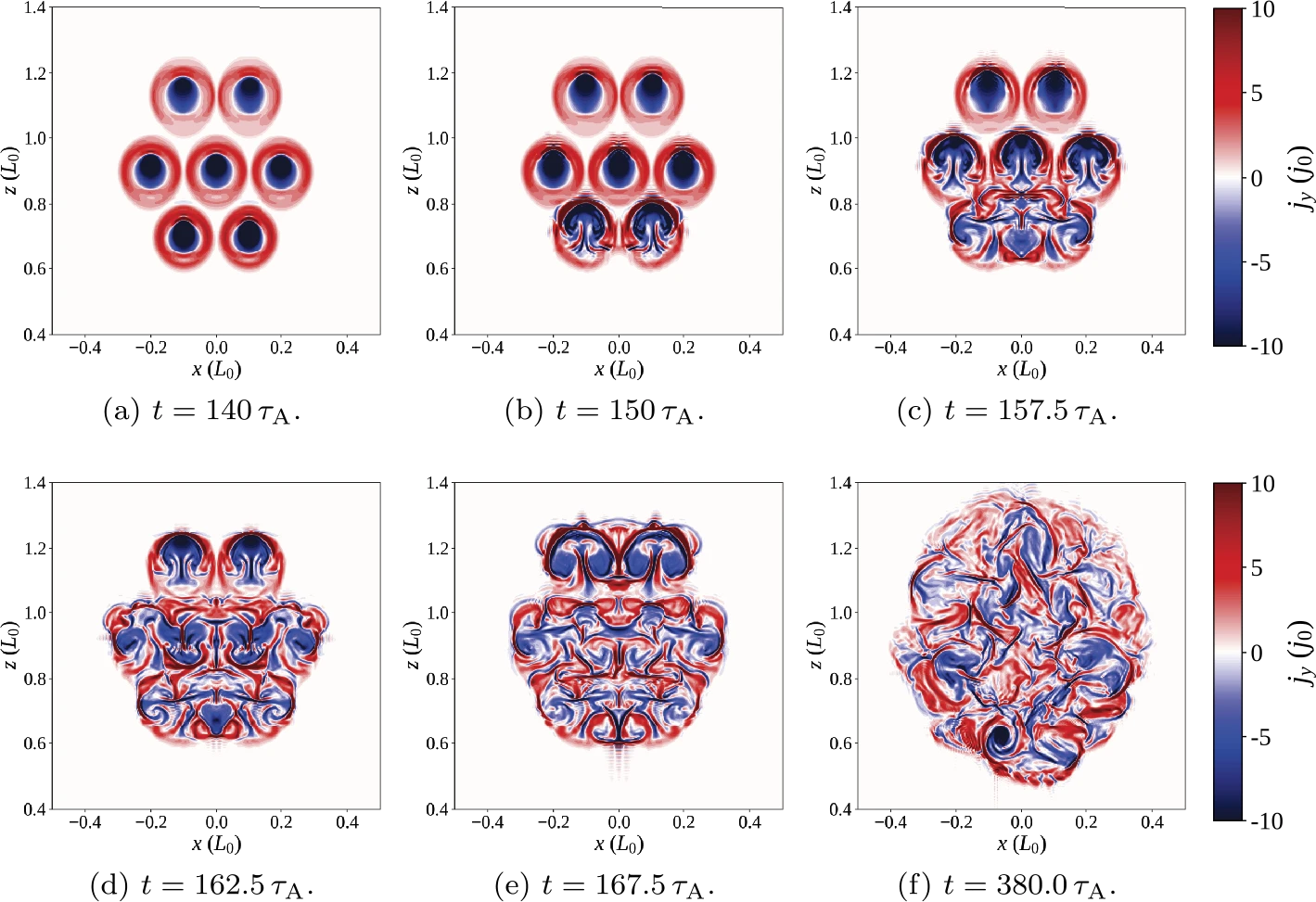}
\caption{Contours of the toroidal currents in a cut above the polarity inversion line, for different times (see details in \cite{Threlfal21})  Reproduced with permission from Threlfal   et al., Solar Physics, {\bf 296}, 120 (2021), Copyright  2021 Springer.}
\label{Thre4}
\end{figure}

Threlfall et al. \cite{Threlfal21} analyzed the evolution of multi-threaded coronal magnetic flux tubes (see Fig. \ref{Thre}), which become unstable when varying the driving velocity of the individual threads.  Assuming that one of the threads is  destabilized, it will very quickly lead the system of threads to a strongly turbulent environment, forming many CoSs of different scales, and it reorganizes the multi-threaded topology into a complex magnetic topology with several multi scale intermittently appearing and disappearing current filaments (Fig. \ref{Thre4}) (see details in Threlfall et al. \cite{Threlfal21} and references therein).

Archontis and Hansteen \cite{Archontis14} reported on the formation of CoSs generated by patchy magnetic reconnection between interacting
magnetic loops as well, using a different initial setup. A three-dimensional magnetohydrodynamic numerical experiment was performed,
where a uniform magnetic flux sheet was injected into a fully developed convective layer. The gradual emergence
of the field into the solar atmosphere results in a network of magnetic loops, which interact dynamically, forming
current layers at their interfaces. They find that these
CoSs are short-lived (from $30\,$s to  minutes), giving rise to bursts of energy in the range $10^{25}$ $10^{27}\,$erg, which basically is  the
microflare range. The CoSs' persistent formation, interaction, and evolution leads to recurrent
emission of fast EUV/X-ray jets and to considerable plasma heating in the active corona.

All attempts to model the non linear coupling and the evolution of the emerged magnetic fields, driven by photospheric and subphotospheric motions,  can be explored with one more tool. Non linear extrapolation of the force free magnetic filed gives an approximate snapshot of the turbulent  state in a coronal active region.

\begin{figure}[ht!]
%\sidesubfloat[]
\includegraphics[width=0.6\columnwidth]{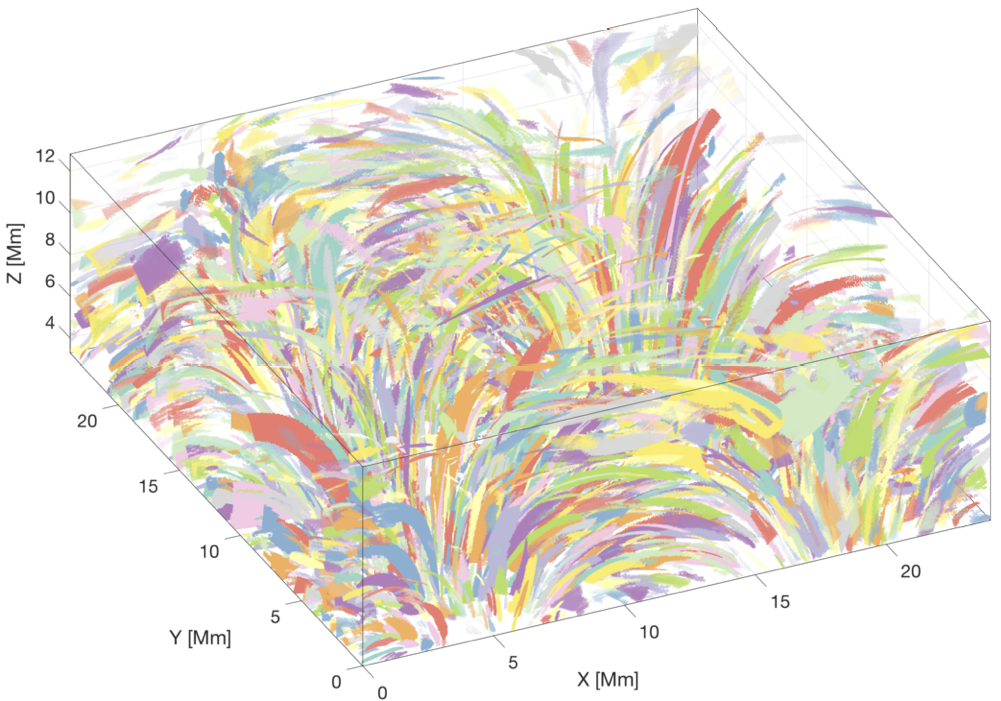}%
		%\label{Turk}}
	\hfill\\
		%\sidesubfloat[]
		{\includegraphics[width=0.6\columnwidth]{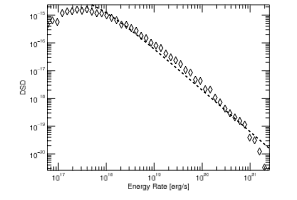}%
		\label{Turk}}\hfill
		%\sidesubfloat[]{\includegraphics[width=0.55%\columnwidth]{f1c}%
		%\label{MHD_t}}\hfill\\
	\caption{ (a) 3D magnetic topology in the solar corona, with 4136 identified CSs; each color represents a different CS. (b) Differential size distribution of the identified CSs' energy rate in logarithmic scale  (diamonds).
A power law fit (dashed) yields the index $\sim  -1.50 \pm 0.02$ (see Kanella and Gudisken \cite{Kanella17} for details.  Reproduced with permission from Kanella and Gudisken, Astronomy and Astrophysics, {\bf 603}, 83 (2017), Copyright  2021 ESO.
			}\label{Kanella}
\end{figure}

Kanella and Gudisken \cite{Kanella17, Kanella18, Kanella19} used a 3D MHD numerical code to simulate the region from the solar convection zone up to the solar corona. The code includes different processes occurring in the convection zone, photosphere, chromosphere, transition region and corona.  The simulated volume starts $2.5\,$Mm below the photosphere and extends $14.3\,$Mm above the photosphere into the corona. They   used periodic boundary conditions in the horizontal $x$â $y$-plane; in the vertical $z$-direction, the upper boundary is open, while the lower boundary is also open, but remains in hydrostatic equilibrium, enabling convective flows to enter and leave the system.  They
incorporated two strongly magnetic regions of opposite polarity, which are connected through a magnetic structure with a loop-like shape. The magnetic field is initially set vertically at the bottom boundary and extrapolated to the whole atmosphere, assuming a
potential field, while a horizontal  field of $100$ Gauss is continuously fed in at the lower boundary, producing random bipolar structures  in the photosphere.  Identifying locations with 3D current sheets, as we have already discussed in section II,  turns out not to be so simple. The CSs in 3D are generally not 2D flat
structures, as the cartoon-like pictures of 2D reconnection would suggest, but they are much more complex. Often, the background current level is higher in places with many current sheets, so it is not easy to separate one current sheet from a cluster of CSs. This is in
some ways similar to the problems experienced by observers,
where the background is causing large obstacles for the interpretation. The method Kamela and Gudisken \cite{Kanella17} have applied to identify CSs is called  ImageJ, used in medical imaging and bio-informatics to perform multi-dimensional image analysis.  Fig.\ \ref{Kanella}(a) shows the 4136 identified CSs in the coronal part of the simulation volume, for a certain time-instant during the simulation.  In Fig. \ref{Kanella}(b), the differential size distribution of the identified CSs' released
energy rate is plotted in logarithmic scales, which obey a power-law scaling with an index $-1.5$. 

%\begin{figure}[h!]
%\centering
%\includegraphics[width=0.70\columnwidth]{f26a.png}\\
%\includegraphics[width=0.70\columnwidth]{f26b.png}
%\includegraphics[width=0.45\columnwidth]{Dtr1.eps}
%\includegraphics[width=0.45\columnwidth]{Dtr2.eps}
%\caption{(a) 3D magnetic topology in the solar %corona, with 4136 identified CSs; each color %represents a different CS. (b) Differential size %distribution of the identified CSs' energy rate in %logarithmic scale  (diamonds).
%A power law fit (dashed) yields the index $\sim  %-1.50 \pm 0.02$ (see Kanella and Gudisken %\cite{Kanella17} for details).}
%\label{Kanella}
%\end{figure}

Einaudi et al. \cite{Einaudi21} discussed the differences between "nanoflares", introduced by Parker, and "elementary events," defined in their article as small-scale spatially and temporally isolated heating events, resulting from the continuous formation and dissipation of field-aligned current sheets within a coronal loop. They presented numerical simulations of the compressible 3D MHD equations. They used two clustering algorithms to investigate the properties of the simulated elementary events: an IDL implementation of a density-based spatial clustering  technique for applications with noise, and their own physical distance clustering algorithm. They identified and tracked elementary heating events in time, and for every event they characterized properties such as density, temperature, volume, aspect ratio, length, thickness, duration, and energy. The energies of the events are in the range of $10^{18}\,-\,10^{21}\,$erg, with durations shorter than $100\,$s. A few events last up to $200\,$s and release energies up to $10^{23}\,$erg. While high temperatures are typically located at the flux tube apex, the currents extend all the way to the footpoints. Hence, a single elementary event cannot  be detected at present. The observed emission is due to the superposition of many elementary events, distributed randomly in space and time within a loop. 

The formation of CSs and other CoSs in the solar atmosphere, driven by the turbulent convection zone, depends on the magnetic  topology above the photosphere (e.g.\ single magnetic flux tube, multi-flux magnetic tubes, , etc), and on the details of the convection zone driver (small scale stochastic motion, turbulent motions, emerging magnetic flux, etc). All articles listed above followed the initial conjecture of   Parker\cite{Parker72} and focused on the formation of randomly distributed small scale CSs inside a volume, which subsequently will provide Ohmic heating in the Corona.  It was assumed, without any proof,  that the current sheets formed and the energy released are very small (``nanoflares'' or ``elementary events'', with energy in the range  $ \approx 10^{20}\, - \,10^{21}\,$erg). In our opinion, this assumption was not supported by the MHD simulations, since the flux tubes, or the multiple flux tubes, or the complex magnetic topologies are filled with CoSs at all scales. The observed explosive phenomena and flares in closed magnetic topologies are also present in the strongly turbulent solar corona  and follow the statistical properties reported \cite{Crosby93} when the turbulent corona is driven by turbulent  photospheric motions. We must emphasize here that the magnetic coupling of the convection zone with the solar atmosphere and the formation of CoSs is a multi-scale process, ranging form $10^{11}\,$cm to a few cm.

\subsubsection{Formation of CoSs during  explosive events associated with large scale reorganizations of the magnetic field in the solar atmosphere}

A 3D study of solar explosions associated with CMEs is also a multiscale problem. Cheung et al. \cite{Cheung19} used a 3D MHD code to capture the large scale of such explosions. Their simulation domain captured the top $7500\,$km of the solar convection zone, and the first $41600\,$km of the overlying solar atmosphere. The initial set up was inspired by the observed evolution of a specific Active Region (AR), but was not intended to model a specific flare of the specific AR. The initial setup consisted of a bipolar sunspot pair, each with a magnetic flux of $3.4 \times 10^{21}\,$Mx. A strongly twisted magnetic bipole with $10^{21}\,$Mx flux  emerged in proximity to one of the pre-existing sunspots. The emergence of the parasitic bipole leads to the creation of a twisted coronal flux rope well before the flare onset. The code of Cheung et al. \cite{Cheung19} is ideal to capture the large scale evolution, but it misses all the physical processes related with the formation of CoSs and their interaction with particles. Therefore,  two more levels of scales are necessary to be included in such an analysis. A hybrid code can capture the formation of CoSs (meso-scales), and a kinetic code can explore the details of the collective dissipation of CoSs and the heating and acceleration of particles.

\begin{figure}[h!]
\centering
\includegraphics[width=0.80\columnwidth]{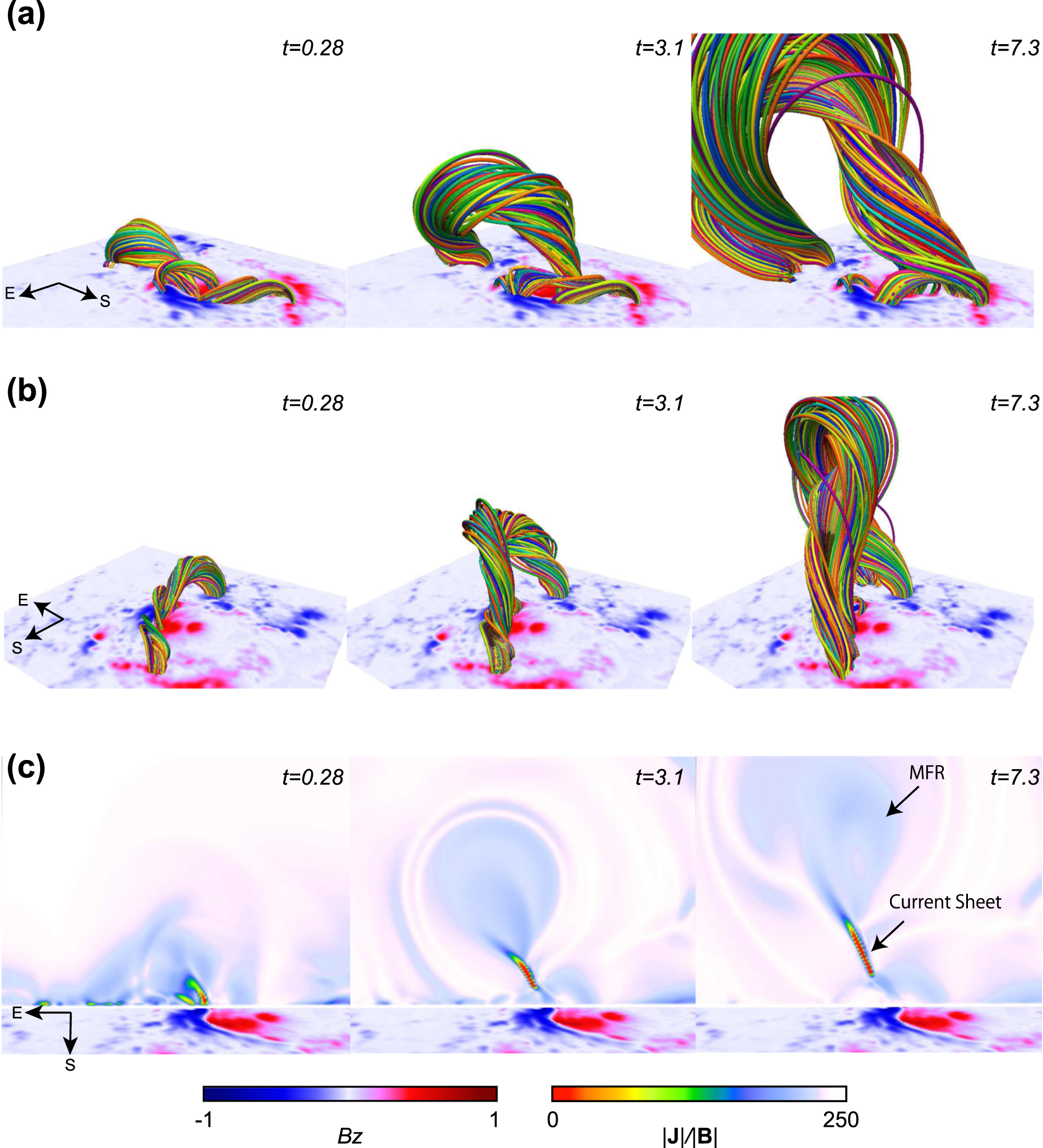}\\

\caption{ Temporal evolution of the formation and dynamics of the eruptive magnetic flux ropes (MFRs). Panels (a) and (b) show the field lines from different viewing angles. E and S stand for east and south. (c) Temporal evolution of $| \mathbf{J}| /| \mathbf{B}| $,  plotted in the $x$â $z$-plane.  Reproduced with permission from Inoue  et al., The Astrophysical Journal, {\bf 867}, 83 (2018), Copyright  2018 AAS}
\label{Inoue}
\end{figure}

Inoue et al. \cite{Inoue18, Inoue23} used a 3D MHD code to analyze  a specific flare/eruption of a specific AR. Their starting point was the nonlinear force free field (NLFFF) extrapolation  \cite{Inoue16} of the magnetogram.  In Fig.\ \ref{Inoue}, the large scale evolution of the eruption is shown. The striking part in this simulation is the presence of thousands of magnetic flux ropes (MFR), which are twisted and evolve, possibly interacting with each other. Based on the analysis presented in the previous subsection, an individual MFR \cite{Galsgaard96} and a collection of MFRs \cite{Threlfal21} can create a dense environment of CoSs along with the evolving MFRs and their interaction. As shown in Fig. \ref{Inoue}, Inoue et al focused their analysis on the formation of the large scale CS in the middle of the huge structure.  The fragmentation of this structure and the formation of smaller scale structures, as we reported earlier, was beyond the resolution of their simulation (see also  He et al. \cite{He20}).

\begin{figure}[h!]
\centering
\includegraphics[width=0.90\columnwidth]{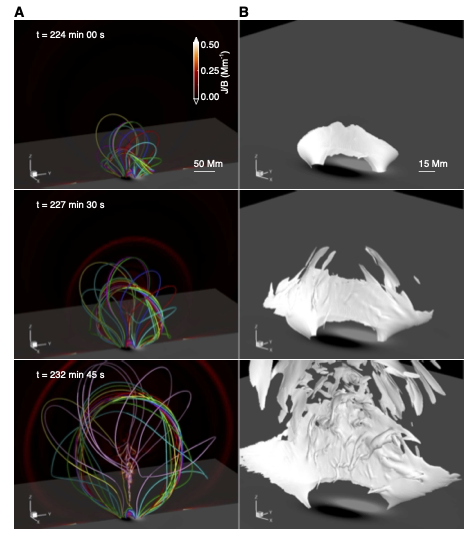}
\caption{Evolution of magnetic field lines and the large scale CS in 3D during the eruption.
(A) The magnetic field lines are shown by thick colored lines, where the colours are
used for a better visualization of the different lines. Note that the MFR is weakly
twisted in its core but highly twisted in its envelope. The bottom surface is shown with the
distribution of the magnetic 
flux. The vertical, transparent slice shows the distribution
of the current density normalized by the magnetic field strength, i.e.\ $J/B$. (B) The CS in the 3D
configuration is shown as iso-surface for $J/B = 0.5\,$Mm$^{-1}$. Reproduced with permission from Jiang  et al., Physics of Fluids, {\bf 33}, 055133 (2021), Copyright  2018 AIP}
\label{Jiang}
\end{figure}

The evolution of a large collection of MFRs was studied with the use of 3D MHD codes recently by several authors (see \cite{He20, Jiang21b} and references therein). Jiang et al. \cite{Jiang21b} used a fully 3D  MHD simulation with high accuracy to follow solar eruptions. They initiated the simulation with a bipolar configuration with no additional special magnetic topology. The bipolar configuration was driven unstable by a photospheric shearing motion. Once the large-scale CoSs and CSs are formed and  reconnection starts, the whole arcade expands explosively, forming fast expanding twisted flux ropes with very strong turbulence in the volume underneath (see Fig.\ \ref{Jiang}). The simplicity and efficiency of their scenario highlights the importance of the magnetic topology driven unstable and the role of the driver. After all, when using a realistic magnetic topology and a turbulent photospheric driver or emerging magnetic field, large scale CoSs very quickly fragment (see more on this in section III) and drive strong turbulence.  

 Several authors \cite{Gordovsky11,  Rappazzo13}, analyzed the stressing of an isolated  flux rope (FR) (which is part of the highly  stressed large scale magnetic topologies shown in Figs. \ref{Inoue} and \ref{Jiang}) at the two ends by large scale  localized photospheric vortical motion, which twists the coronal field lines, and the resulting current fragmentation reaches again the state of  strong turbulence, as discussed earlier. 

%\begin{figure}[h]
%\centering
%\includegraphics[width=0.80\columnwidth]{kinkra.eps}
%\includegraphics[width=0.40\columnwidth]{Mor2.eps}\\

%\includegraphics[width=0.45\columnwidth]{Dtr1.eps}
%\includegraphics[width=0.45\columnwidth]{Dtr2.eps}
%\caption{Isosurfaces of the squared current $j^2$ at a time after the start %of the kink instability \cite{Rappazzo13}. \heinzC{The figure is not %mentioned in the text.} }
%\label{Kink}
%\end{figure}

As we mentioned already in Sect.\ 3,  magnetic flux emergence and the subsequent eruptions, with the formation of jets,  was studied in many articles \cite{Heyvaerts77,Archontis04,Archontis05,Galsgaard05, Archontis12a, Archontis12b, Moreno-Insertis13,Raouafi16, Wyper16, Wyper17}. Most of the numerical studies stopped their analysis at the formation of a large scale current sheet through the interaction of the emerging flux tube with the ambient magnetic field of the solar atmosphere.

We can then conclude that during large scale magnetic explosions, the fragmentation of formed large scale CSs, and the formation of CoSs through the stresses resulting from eruptions inside MFRs reported above, are extremely important for the analysis of the heating and acceleration of the coronal plasma during eruptions and large scale reorganization of the magnetic field of an active region. The fact that  solar flares and the associated CMEs were modeled so far as 2D structures  with a monolithic large scale CS as the main source of the energy released (called by many researchers the \textbf{standard flare}) has misled the analysis of the observed data resulting from the  heating and acceleration of particles during solar explosions.

%\subsection{Nanoflares and turbulent reconnection in the quite corona}

\section{Formation and evolution of  coherent structures  and self organized criticality}

\begin{figure}[h]
\centering
\includegraphics[width=0.90\columnwidth]{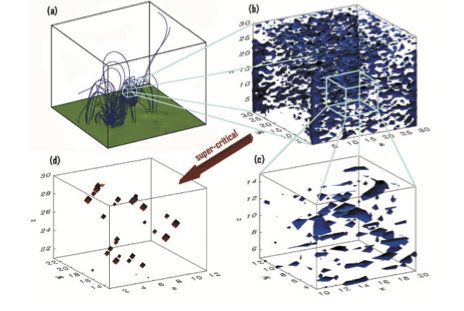}
\caption{ (a)  Force-free magnetic field lines, extrapolated from the convection zone into the corona. (b) CoSs in a sub-volume of a coronal active region. (c) Same as (b), but zoomed. (d) Spatial distribution of the reconnecting CSs inside a sub-volume of the complex active region. Reproduced with permission from Vlahos  et al., The Astrophysical Journal, {\bf 608}, 540 (2004), Copyright  2004 AAS }
\label{Multi}
\end{figure}

We have repeatedly stated in this review that many space and astrophysical systems have a  turbulent driver as a source of their strong turbulence, e.g.\ the convection zone acts as a driver for the magnetic field extending from the convection zone into  the solar atmosphere and the solar wind,  solar wind turbulence acts as the driver in the vicinity of the Earths bow shock, the magnetosheath, and the magnetotail, etc. In the previous section, we focused our analysis on the magnetic coupling of the convection zone with the solar atmosphere (see Fig.\ \ref{Multi}), utilizing mainly 3D MHD codes to concentrate on the evolution of the driven system at the large scales  \cite{Vlahos16a}.   

In Fig.\ \ref{Multi},  we present a scenario to reconstruct the formation of CoSs in an active region, using a simple cartoon. The starting point is a 3D large scale magnetic topology, which can be estimated by force-free extrapolation of the photospheric magnetic field.  The more intense magnetic structures are forming active regions (ARs) (see Fig.\ \ref{Multi}a). As discussed extensively in Sec.\ \ref{CoS_Shocks}, the formation of CoSs inside an AR is shown in Fig.\ \ref{Multi}b. Only a small number of CoSs are forming CSs inside an AR, and from them an again small fraction will reconnect (see Fig.\ \ref{Multi}c, d). This is a multi scale system, extending from  $100$ - $1000\,$Mm in the initial box (Fig.\ \ref{Multi}(a))  down to a few cm's at the scale of the reconnecting CSs (Fig.\ \ref{Multi}d).

The existing numerical tools cannot handle the multi scale coupling of the convection zone with the solar atmosphere, so we are searching for other tools to explore the formation of CoSs in the solar atmosphere.

In this section, we will concentrate on numerical tools that are used extensively in the analysis of complex systems (e.g.\ systems far away from equilibrium, like strong turbulence, or systems comprised of a large number of nonlinearly interacting sub-systems, like a collection of CoSs in a turbulent plasma). The most popular numerical tool used to analyze complex systems is the Cellular Automaton (CA) model \cite{ChopardB}. The set up of the CA depends strongly on a qualitative analysis of the physical system under study, which guides the definition of the rules of the CA. The success  of a CA model is assessed by the direct
comparison with data and results from MHD simulations. So complex systems, like turbulent plasmas, can be explored by CA models (on the global astrophysical scales), and by MHD or kinetic simulations on the intermediate and the local scales. Also, MHD and kinetic simulations can serve as tools for defining the rules of a CA.

The existence of power laws in the frequency distributions of the
explosive solar activity (see Fig. \ref{fig:2})  may suggest that explosions are a self-organization
phenomenon in ARs. Lu and Hamilton \cite{Lu91} (LH91) were  the first to
realize that ARs may be in a self-organized critical
state, and they proposed that explosions  ultimately are caused by small magnetic
perturbations $(\delta B)$ ({\bf loading}), which gradually force a CS to reconnect  when a {\bf local critical threshold} is passed. The local fragmentation of the reconnecting CSs (see section III) causes a re-organization of
the unstable magnetic topology, which may cause
{\bf avalanches}  of CSs at all scales to reconnect and to release energy 
 (nano-flares, micro-flares, flares) (the basic ideas of SOC were initially proposed by Bak et al. \cite{Bak87}, thirty five years ago). The LH91 model opened
the way for a series of similar models developed during the  last twenty five years (see the reviews by \cite{Charbonneau01, Aschwanden11, Aschwanden16, Vlahos16a}).

There are many ways to develop  CA models to represent  SOC \cite{Aschwanden11}, one of them based its rules on the MHD equations \cite{Isliker00, Isliker01}. The proposed  set-up can be superimposed
onto each classical solar flare CA model,  making the
latter interpretable in a MHD-consistent way ({\it classical}
CA models here means the LH91 model\cite{Lu91}  and
its modifications, which are based on the sand-pile analogy \cite{Vlahos95, Georgoulis96, Georgoulis98a}).
The set-up thus specifies the physical interpretation of the
grid-variables and allows the derivation of quantities such as
currents etc. It does not interfere with the dynamics of the CA
(unless wished): loading, redistributing (bursting), and the
appearance of avalanches and Self-Organized Criticality (SOC), if the latter are implied by the evolution rules, remain unchanged.
The result is therefore still a CA model, with all the advantages
of CAs, namely that they are fast, that they model large spatial
regions (and large events), and therewith that they yield good
statistics. Since the set-up introduces all the relevant physical
variables into the context of a CA model, it automatically
leads to a better physical understanding of the CA models. It
reveals which relevant plasma processes and in what form are
actually implemented, and what the global flare scenario is the CA
models imply. All this is more or less hidden otherwise in the
abstract evolution rules. It leads also to the possibility to
change the CA models (the rules) at the guide-line of MHD, if this
should become desirable. Not least, the set-up opens a way for
further comparison of the CA models to observations.

The specifications the set-up meets are: The vector $\mb{A}_{ijk}$ at the grid sites $\mb{x}_{ijk}$ denotes the local
vector-field, $\mb{A}(\mb{x}_{ijk})$. Note that this was not
specified in the classical CA models. Lu et al. \cite{Lu93} for instance
discussed this point: it might also have been thought of as a mean
local field, i.e. the average over an elementary cell in the grid.

Guided by the idea that one wants to assure $\nabla \cdot \mb{B} = 0$ for
the magnetic field $\mb{B}$, which is most easily achieved by
having the vector-potential $\mb{A}$ as the primary variable and
letting $\mb{B}$ be the corresponding derivative of $\mb{A}$
($\mb{B} = \nabla \times \mb{A}$), it is furthermore assumed that the
grid variable $\mb{A}$ of the CA model is identical with the
vector-potential.

The remaining and actually most basic problem then is to find an
adequate way to calculate derivatives in the grid. In general, CA
models assume that the grid-spacing is finite, which also holds
for the CA model of \cite{Lu91} (as shown in detail by
\cite{Isliker98}), so that the most straightforward way of replacing
differential expressions with difference expressions is not
adequate. Consequently, one has to find a way of continuing the
vector-field into the space in-between the grid-sites, which will
allow to calculate derivatives. For this purpose, \citet{Isliker00, Isliker01} proposed to use spline interpolation, where the 3D interpolation is
performed as three subsequent 1D interpolations in the three
spatial directions \cite{Press92}. For the 1D splines, 
natural boundaries are assumed (the second derivatives are zero at the
boundaries).

With the help of this interpolation, the magnetic field $\mb{B}$
and the current $\mb{J}$ are calculated as derivatives of $\mb{A}$, according to the MHD prescription:
\begin{equation}
\mb{B} = \nabla \times \mb{A},
\end{equation}
\begin{equation}
\mb{J} = \frac{c}{4\pi} \, \nabla \times \mb{B} .
\end{equation}

According to MHD, the electric field is given by Ohm's law, $\mb{E} = \eta \mb{J} - \frac{1}{c} \mb{v} \times \mb{B}$, with $\eta$
the diffusivity and $\mb{v}$ the fluid velocity. Since the
classical CA models use no velocity-field, the set-up can yield
only the resistive part,
\begin{equation}\label{res}
\mb{E} = \eta \mb{J}.
\end{equation}
In applications such as to solar explosions, where the interest is in
current dissipation events, i.e.\ in events where $\eta$ and $\mb{J}$ are strongly increased, Eq. \ref{res} can be expected to be a
good approximation to the electric field. Theoretically, the
convective term in Ohm's law would in general yield just a
low-intensity, background electric field.

Eq.\ (\ref{res}) needs to be supplemented with a specification of the
diffusivity $\eta$: \citet{Isliker98} have shown that in the classical
CA models the diffusivity adopts the values $\eta =1$ at the
unstable (bursting) sites, and $\eta =0$ everywhere else. This
specifies Eq.\ (\ref{res}) completely.
The set-up of Isliker et al. \cite{Isliker00, Isliker01}   for classical solar flare CA
models yields, among others, consistency with Maxwell's equations
(e.g.\ divergence-free magnetic field), and availability of
secondary variables such as currents and electric fields in
accordance with MHD. The main aim in \citet{Isliker00, Isliker01} with the introduced set-up was
to demonstrate that the set-up truly extends the classical CA
models and makes them richer in the sense that they contain much
more physical information. The main features they revealed about the classical CA
models, extended with their set-up, are:

\noindent {\bf 1.\ Large-scale organization of the
vector-potential and the magnetic field:}
The field topology during SOC state is bound to characteristic
large-scale structures which span the whole grid, very pronounced
for the primary grid variable, the vector-potential, but also for
the magnetic field. Bursts and flares are just slight
disturbances propagating  over the large-scale structures, which
are always maintained, also in the largest events.

\noindent {\bf 2.\ Increased current at unstable grid-sites:}
Unstable sites are characterized by an enhanced current, which is
reduced after a burst has taken place, as a result of which the
current at a grid-site in the neighbourhood may be increased.

\begin{figure}[ht!]
\centering
{\includegraphics[width=0.70\columnwidth]{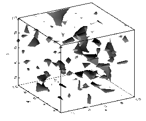}}

\caption{Three dimensional isosurfaces of the electric current
density, as yielded by the CA model of  \citet{Isliker00}. Reproduced with permission from Isliker  et al., Astronomy and Astrophysics, {\bf 363}, 1068 (2000), Copyright  2000 ESO}
\label{CS1}
\end{figure}

\noindent {\bf 3.\ Availability of the electric field:} The
electric field is calculated with the resistive part of Ohm's law,
which can be expected to be a good approximation in applications
where the interest is in current-dissipation events, e.g.\ in the
case of solar flares.

\noindent {\bf 4.\ Energy release in terms of Ohmic dissipation:}
\citet{Isliker00, Isliker01} also replaced the some-what {\it ad hoc formula} in the classical CA models to
estimate the energy released in a burst
with the expression for Ohmic dissipation in terms of the current.
The distributions yielded in this way are very similar to the ones
based on the ad hoc formula, so that the results of the CA models
remain basically unchanged.

\noindent {\bf 5.\ CA as models for current dissipations:} As a
consequence of point 2 and 4 in this list, and of the fact that
there is an  approximate linear relation between the current and
the stress measure of the classical CAs, one can conclude that the {\it
extended} CA models can be considered as models for energy release
through current dissipation.

It is rather interesting to compare the 3D isosurfaces shown in  Fig.\ \ref{CS1} for the electric field generated by the CA model of \citet{Isliker00, Isliker01} with the 3D MHD simulations reported in \citet{Galsgaard96} and shown here in Fig.\ \ref{Dtr}. 
It is remarkable to note
the appearance of non-steady current surfaces in the MHD model, as it is the case
in the CA model in Fig.~\ref{CS1}.

Fragos et al. \cite{Fragos04} used ``magnetograms" generated with the percolation method and applied a linear extrapolation to search for the statistical properties of the reconstructed coronal AR. \citet{Moraitis16} did the same, using observational magnetograms, and they noticed that  the re-organization of the magnetic fields is a potential way to identify reconnecting CSs in the coronal part of an AR.
Dimitropoulou et al. \cite{Dimitropoulou11,  Dimitropoulou13} used a series of observed magnetogarms to drive the SOC model proposed by Isliker et al. \cite{Isliker00, Isliker01}. They obtained robust power laws in the distribution functions of the modeled flaring events, with scaling law indices that agree well with the observations.

\begin{figure}[ht!]
\centering
\resizebox{0.50\columnwidth}{!}{\includegraphics{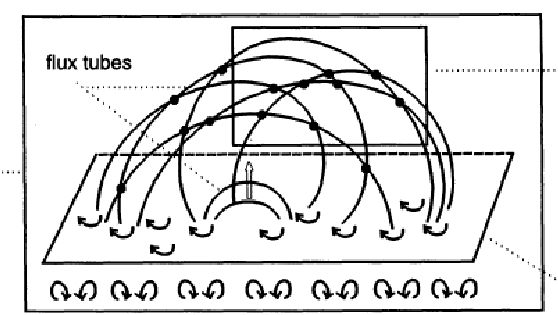}}
\resizebox{0.45\columnwidth}{!}{\includegraphics{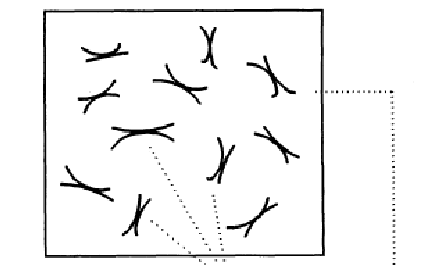}}
%\resizebox{0.30\columnwidth}{!}{
%\includegraphics{TurbE.eps}}\resizebox{0.30\columnwidth}{!}{
%\includegraphics{curturb.eps}
%}
\caption{Formation of CSs developed intermittently at
random positions. Reproduced with permission from Anastasiadis and Vlahos, The Astrophysical Journal, {\bf 428}, 819 (1994), Copyright  1994 AAS .
}
\label{vl1}
\end{figure}

Models along these lines have been proposed
\cite{Vlahos94a, Anastasiadis94} (see also Fig.~\ref{vl1})  in the mid
80's and the beginning of the 90's and remained undeveloped due to
the lack of tools for the global analysis of active regions
till recently.
The nonlinear coupling of the turbulent convection zone with ARs, as well as the consistency of the related results obtained by the SOC theory with those from  turbulence simulations, have been studied intensively by several authors \cite{Uritsky07}. Uritsky et al. \cite{Uritsky12}  examined in depth the question of the relation of SOC with turbulence in the solar corona and agreed  with the suggestion made  by Dahlburg et al. \cite{Dahlburg05} that  reconnecting CSs and their \
{\bf fragmentation can serve as the driver for  avalanches in the SOC scenario}. Uritsky \& Devila \cite{Uritsky12} also suggested, by studying an AR in a quiescent non-flaring period, that (1) there is  formation of non-potential magnetic structures with complex polarity separation lines inside the active region, and (2) there are statistical signatures of canceling bipolar magnetic structures coinciding with flaring activity in the active region. Each of these effects can give rise to an unstable magnetic configuration, acting as an energy source for coronal dissipation and heating.
The development of a parallel use of models based on complexity theory and of well established 3D MHD or kinetic codes is the only way to explore the interplay between global and local scales in turbulent systems.

The tools used in this section focused on  reconnecting CSs and cannot capture all types of  CoSs in the driven turbulent system present in the solar atmosphere. Keeping track of CoSs  that are not leading to reconnection, but still dissipate energy collectively in a large volume, is an open project for future models that will utilize tools of complexity theory.

\section{Discussion ans Summary}

The formation of CoSs  in 3D strongly turbulent magnetized plasma remains till today an open problem for the analysis of space, astrophysical and laboratory plasmas. The main obstacle remains the multi-scale characteristics,  with the dynamic evolution taking place on all scales. The list of intermittently appearing CoSs inside 3D strongly turbulent magnetised plasmas is long (CSs (non-reconnecting and reconnecting), magnetic filaments, large amplitude magnetic disturbances, vortices, shocklets, etc) and expanding. CSs and their reconnection are the best studied CoSs, but so far the vast majority of studies separated CSs and their evolution from the rest of the CoSs and the turbulent environment where they were formed. The other shortcoming in the analysis of CSs till today is the fact that their evolution was analyzed with  analytical and numerical tools in 2D magnetic topologies.  The other well known large scale magnetic discontinuities that were isolated from CoSs and a strongly turbulent environment in their vicinity are large scale shocks. Isolating CSs and shocks in their analysis from 3D magnetic topologies and the strongly turbulent environments, in which they are formed and evolve, can lead  to an erroneous answer to the question ``Who really needs turbulence?".

In this review, we have addressed two important questions: (1) How CoSs are formed? We presented three different ways for the formation: (a) The formation of CoSs (with special emphasis on CSs) where 3D strong turbulence is present. Common space,  astrophysical, and laboratory plasmas, where strong turbulence is present, are the solar atmosphere, the solar wind, astrophysical jets, edge localised mode turbulence in TOKAMAKS, etc. (b) The fragmentation of large scale CSs, appearing mostly in explosive phenomena in the solar atmosphere, in the magnetotail under the influence of the solar wind, and in the magnetopause under the influence of the magnetosheath. (c) The interaction of large scale shocks with  strong turbulence upstream and downstream of the shock, as appearing in the interaction of the Earth's bow shock with the turbulent  solar wind, at the termination shock in the heliosphere, and at super nova remnants.
(2) How strong turbulence is excited in astrophysical settings and in laboratory plasmas.  We have chosen in this review to address mainly the excitation of strong turbulence by the convection zone, by turbulent flows,  and by the fragmentation of large scale CSs formed during coronal explosions and in the  magnetotail.

Let us now summarize the main points in this review.

\begin{itemize}
    \item CoSs are formed in strongly turbulent 3D magnetized plasmas.
    \item CSs formed inside a 3D strongly turbulent plasma cannot be analyzed as isolated 2D periodic structures. 
    \item Only a small fraction of the CSs formed inside a 3D strongly turbulent plasma reconnect. Therefore, magnetic reconnection  dominates the acceleration of the small fraction of high energy particles in the tail of the particles' energy  distribution. The collective interaction of the non reconnecting CSs at all scales and  with other CoSs (large scale magnetic disturbances, etc.) may play a crucial role in the heating of the ambient plasma and the overall  dissipation of energy. 
    \item The methods developed so far to search in observed data for 3D CoSs in strongly turbulent plasmas are not sufficient to capture their statistical properties, since they have been based mainly on the 2D modeling of the characteristics of CoSs. 
    \item The formation of large scale CSs and their subsequent fragmentation inside a turbulent plasma gives rise to clusters of CoSs in the vicinity of the CSs and contributes to the development of smaller scale activity inside the strongly turbulent system. 
    \item Karimabadi et al. \cite{Karimabadi2014}  and Gro\v selj et al. \cite{Groselj19} emphasize that the motion of CoSs generates waves that are emitted into the ambient plasma in the form of highly oblique compressional and Alfven modes, as well as large amplitude magnetic disturbances.  This indicates that strong turbulence will in general consist of CoSs and waves, therefore "weak" and strong turbulence  co-exist in the multiscale evolution of a strongly turbulent plasma.
    \item Large scale magnetic disturbances and CoSs in fully developed turbulence follow a mono-fractal or multi-fractal scaling, both in space and astrophysical plasmas. This  strongly affects the interaction of particles with CoSs.
    \item Large scale shock waves, like CSs, never appear in isolation. They are formed in the presence of turbulent flows upstream and downstream. The presence of CoSs in the vicinity of a shock solves a number of open problems through the amplification of the CoSs near the shock discontinuity.     
   \item Unfortunately, the multi-scale character and the complexity of the micro-physics present in   evolving large scale shocks cannot be explored by current numerical simulations, and neither can they follow the evolution  for long times. This is the main reason why we  are missing many important details  on the evolution of CoSs and their role in the heating and acceleration of particles so far.
   \item The magnetic coupling of the turbulent convection zone with the solar atmosphere has many avenues for forming CoSs and for exciting strong turbulence in the solar corona and the solar wind. We have emphasized two of them in this review, (1)  magnetic foot-point shuffling of otherwise stable magnetic flux ropes, and magnetic field emerging into a complex magnetic topology, (2) explosive evolution of large scale magnetic structures ( loss of equilibrium, triggered e.g.\ by emerging magnetic flux). 
   \item The multi scale character of CoSs inside a strongly turbulent plasma cannot be addressed with the well known numerical tools (MHD codes, hybrid codes, and particle in cell codes), since they capture only a part of the plasma evolution and cannot realize the self consistent coupling of all scales, and moreover  they neglect the fact that most natural systems are open. 
   \item The use of tools borrowed from complexity theory together with the 3D numerical codes mentioned above can be the solution for addressing the 3D  nature of CoSs appearing inside a strongly turbulent plasma.   
\end{itemize}

The statistical properties of 3D CoSs inside a strongly turbulent plasma are an  interesting and important piece of information when analyzing the transport properties of charged particles in the complex topologies of CoSs. The energy dissipation of CoSs needs more careful analysis, and this problem is currently open and needs a separate review for its exposition.  

\begin{acknowledgements}
We would like to thank our colleagues  Peter Cargill, Tassos Anastsiadis, Manolis Georgoulis, Vasilis Archontis, Rene Kluiving, Marco Onofri, Fabio Lepreti,  Tasos Fragos,  and  Nikos Sioulas, for many interesting and constructive discussions over several years on the topics addressed in this review.   
\end{acknowledgements}

\bibliography{vlahosturb2}
% Produces the bibliography via BibTeX.

\end{document}